\documentstyle[preprint,float,aps,psfig]{revtex} 
\tightenlines
\begin{document}

\title{Charge-Symmetry Violation in Pion Scattering from
        Three-Body Nuclei}

\author{A. E. Kudryavtsev\thanks{kudryavt@heron.itep.ru}
    and V. E. Tarasov\thanks{tarasov@vitep5.itep.ru}}
\address{Institute for Theoretical and Experimental Physics \\
         25 Bolshaya Cheremushkinskaya Street, Moscow 117259,
         Russia}

\author{B. L. Berman\thanks{berman@gwu.edu},
        W. J. Briscoe\thanks{briscoe@gwu.edu},
        K. S. Dhuga\thanks{dhuga@gwu.edu}, and
        I. I. Strakovsky\thanks{igor@gwu.edu}}
\address{Center for Nuclear Studies, Department of Physics \\
         The George Washington University, Washington, D.C.
         20052}

\draft
\date{\today}
\maketitle

\begin{abstract}
We discuss the experimental and theoretical status of 
charge-symmetry violation (CSV) in the elastic scattering 
of $\pi ^+$ and $\pi ^-$ on $^3H$ and $^3He$.  Analysis 
of the experimental data for the ratios $r_1$, $r_2$, and 
$R$ at $T_{\pi}$ = 142, 180, 220, and 256~MeV provides 
evidence for the presence of CSV.  We describe pion 
scattering from the three-nucleon system in terms of 
single- and double-scattering amplitudes.  External and 
internal Coulomb interactions as well as the $\Delta 
_{33}$-mass splitting are taken into account as sources 
of CSV.  Reasonable agreement between our theoretical 
calculations and the experimental data is obtained for $T
_{\pi}$ = 180, 220, and 256~MeV.  For these energies, it 
is found that the $\Delta _{33}$-mass splitting and the 
internal Coulomb interaction are the most important 
contributions for CSV in the three-nucleon system.  The 
CSV effects are rather sensitive to the choice of 
pion-nuclear scattering mechanisms, but at the same time, 
our theoretical predictions are much less sensitive to 
the choice of the nuclear wave function.  It is found, 
however, that data for $r_2$ and $R$ at $T_{\pi}$ = 142
~MeV do not agree with the predictions of our model, 
which may indicate that there are additional mechanisms 
for CSV which are important only at lower energies.
\end{abstract}


\narrowtext
\section{Introduction}
\label{sec:intro}

The issue of charge symmetry violation (CSV) is 
fundamental to our understanding of hadronic interactions, 
and many experimental and theoretical studies have 
addressed this issue (see review \cite{mill}).  In the 
framework of QCD, CSV arises from the mass difference 
between the $u$ and $d$ quarks.  The other principal 
cause for CSV comes from the electromagnetic interaction.

Weinberg \cite{wein} pointed out that the effective 
chiral $\pi N$ Lagrangian, coming from QCD, contains a 
term which violates charge symmetry (see also, a recent 
review by Meissner \cite{meis}).  Thus, not only 
are there kinematic reasons for CSV due to the mass 
differences within baryon multiplets, but direct CSV 
effects should exist as well.  Recently, Gashi 
\textit{et al.} \cite{matt} analyzed low-energy $\pi N$ 
scattering data, and found some indications for direct 
CSV effects in the strong-interaction sector.

Another way to study CSV is through the pion-nuclear 
interaction in the lightest nuclei, particularly via
isomirror elastic scattering.  For the deuteron case, 
the $\pi ^+ d$ cross section is compared with that for 
$\pi ^- d$ in Masterson \textit{et al.} \cite{ma82} and 
Baru \textit{et al.} \cite{baru}, but only small 
differences are found.

For pion elastic scattering from $^3H$ and $^3He$, one 
can consider three mirror ratios formed from the 
differential cross sections:
\begin{eqnarray}
r_1 = \frac
{d \sigma /d \Omega ({\pi ^+}{^3 H})}
{d \sigma /d \Omega ({\pi ^-}{^3 He})},
\phantom{xxxxxxxx}
\nonumber \\
r_2 = \frac
{d \sigma /d \Omega ({\pi ^-}{^3 H})}
{d \sigma /d \Omega ({\pi ^+}{^3 He})},
~{\rm and}~
\phantom{xxxx}
\nonumber \\
R = r_1\cdot r_2, 
\phantom{xxxxxxxxxxxxxxxx}
\label{1}\end{eqnarray}
where $R$ is refereed to as the ``superratio"
\cite{ne90}.  The ${\pi ^+}{^3H}$ and ${\pi ^-}{^3He}$ 
scattering cross sections are isomirror ones, as are 
${\pi ^-}{^3H}$ and ${\pi ^+}{^3He}$.  If charge 
symmetry were conserved, all three ratios would be 
equal to unity.  Of course, the Coulomb interaction is 
not charge symmetric and has to be taken into account. 

The experimental study of these ratios has been 
concentrated on large-angle scattering because the 
Coulomb interaction, which intrinsically violates charge 
symmetry, makes a significant contribution in the 
forward-scattering region.  In a series of LAMPF 
experiments \cite{ne90,pi91,dh96,ma00}, the ratios 
(\ref{1}) were measured in the range of the $\Delta 
_{33} (1232)$ $\pi N$ resonance.  Significant deviation 
(several standard deviations) from unity (up to 
$\sim$20\%) was observed for $r_2$ and $R$ in the 
angular range outside of the Coulomb cone, 
$\theta\geq 30^{\circ}$.  In addition, strong angular 
dependence of both $r_2$ and $R$ was observed in the 
angular interval between $\theta \simeq 60^{\circ}$ and 
$90^{\circ}$.  These CSV effects are much more 
pronounced for $^3H$ and $^3He$ than for $^2H$, where 
the observed asymmetry $A_{\pi}$ 
\footnote{Historically, the CSV experimental data for 
          $\pi ^{\pm} d$ elastic scattering have been 
          evaluated in terms of the asymmetry $A_{\pi}$:
\begin{eqnarray}
A_{\pi} = \frac
{d \sigma /d \Omega ( \pi ^- d) - d \sigma 
/d \Omega ( \pi ^+ d)}{d \sigma /d \Omega 
( \pi ^- d) + d \sigma /d \Omega ( \pi ^+ 
d)}.
\nonumber\end{eqnarray}
          If we define $r_d$ by analogy with the ratios 
          (\ref{1}), we get
\begin{eqnarray}
r_d = \frac{d \sigma /d \Omega ( \pi ^- d)}
{d \sigma /d \Omega ( \pi ^+ d)} = 1 + 
\epsilon ,
\nonumber\end{eqnarray}
          and for small $\epsilon$ we have $A_{\pi} = 
          \epsilon /2$.  Thus, $r_d \simeq 1 + 2 A
          _{\pi}$.}
is not nearly as large: it is only about 2\% 
(see, for example, Ref.~\cite{ma82}).  
In addition, inelastic-scattering data for the ratios 
(\ref{1}) at excitation energy below 20~MeV have been 
reported in Ref.~\cite{barry}, and elastic differential 
cross sections at back angles in Ref.~\cite{ma95}.

Thus, CSV effects manifest themselves clearly in the 
three-nucleon system, even if they are very small in 
the deuteron case.  For this reason, the main goal of 
our theoretical analysis is the elucidation of the 
mechanisms that enhance CSV in the three-nucleon 
system.  In other words, can the observed enhancement 
be due only to the well-known mass differences of the 
hadron multiplets together with the Coulomb interaction?  
Previous studies have indicated that the main reason 
for CSV in the case of the deuteron is the mass 
difference of the charge $\Delta _{33}(1232)$-isobar 
states.  The influence of this effect on the scattering 
amplitude of the $\pi d$ elastic scattering was 
discussed in the papers by Masterson \textit{et al.} 
\cite{ma82} and Baru \textit{et al.} \cite{baru} for
single and single-with-double scattering, respectively.  
Naturally, as the number of multiple-scattering diagrams 
increases with an increasing number of nucleons, one 
expects the effect of the $\Delta _{33}$-mass splitting 
to be more prominent for $^3H$ and $^3He$ than for $^2H$.

In addition to the interaction between charged pions and 
nuclei due to the external Coulomb force, there is an 
internal Coulomb interaction due to the difference in the 
wave functions (WFs) of $^3H$ and $^3He$.  (Note that in 
terms of the strong interaction, there is no difference 
between the WFs of $^3H$ and $^3He$.)  One difference 
in these WFs arises from the additional Coulomb 
repulsion between the two protons in $^3He$, which is 
not present in $^3H$.  

In Ref.~\cite{kim}, the difference in the structure of 
$^3H$ and $^3He$ has been described by a first order 
optical potential in the pion-nuclear interaction in 
which the $^3H$ and $^3He$ charge and magnetic form 
factors extracted from experimental data on elastic 
electron scattering from $^3H$ and $^3He$ were used.  A 
more detailed analysis of pion-$^3H/^3He$ elastic and 
inelastic scattering was reported in Ref.~\cite{kam}, 
where an optical single-scattering pion-nuclear potential, 
calculated at a microscopic level with a realistic 
three-nucleon wave function, was used.  No reasonable 
description of the CSV effects was achieved when the 
calculations were performed with isospin-symmetric wave 
functions for both nuclei.  A different approach was 
used in Ref.~\cite{GG}, where the difference in the WFs 
of $^3H$ and $^3He$ was taken into account explicitly.  
In all of these approaches \cite{kim,kam,GG}, the 
difference in the structure of the three-nucleon 
system results in a sharp enhancement of the ratios 
$r_2$ and $R$ in the angular range 
$60^{\circ}\leq\theta\leq 90^{\circ}$.

One of the reasons for this behavior of the angular 
distribution is that the elastic differential cross 
section in the single-scattering approach contains a 
significant contribution at small scattering angles 
and has a minimum at $\theta\simeq 90^{\circ}$ which is 
related to the dip in the $\pi N$ non-spin-flip 
amplitude.  Therefore, even a small contribution to the 
interaction that violates charge symmetry ({\it {e.g.}}, 
the Coulomb interaction) can produce an enhanced effect.  
The influence of the $\Delta _{33}$-mass splitting on 
the observed CSV effect might also result in a large 
effect in this angular range.  

Moreover, to understand the angular distribution in 
detail, one must look beyond the single-scattering 
approach to the $\pi N$ interaction.  We therefore
examine the contributions of both single- and 
double-$\pi N$ scattering to the pion-nuclear 
scattering amplitude following techniques developed
in Ref.~\cite{baru}.  We then take CSV effects into 
account to obtain expressions for the ratios 
(\ref{1}), which then are compared with the 
experimental data \cite{ne90,pi91,dh96,ma00}.

An analysis of the experimental status for the 
ratios (\ref{1}) at energies spanning the $\Delta 
_{33}$ resonance is given in Section~\ref{sec:expt}.  
In Section~\ref{sec:theor1}, we explain how the 
basic ingredients of the scattering amplitude and 
the constraints of single and double scattering are 
combined for $\pi ^3H$ and $\pi ^3He$ elastic 
scattering.  In Section~\ref{sec:theor2}, we derive 
expressions for the scattering amplitudes, taking 
into account all three effects responsible for CSV.  
In Section~\ref{sec:theor3}, we discuss the influence 
of these factors on the ratios (\ref{1}) to show the 
effect of the individual CSV factors, and we compare 
the results of our calculations with the data.  In 
Section~\ref{sec:add}, we discuss some related 
issues associated with CSV effects in other nuclei 
over broader ranges in energy and scattering angle.  
In Section~\ref{sec:conc}, we summarize our findings.

\section{Analysis of the Experimental Status} 
\label{sec:expt}

CSV in pion-nucleus scattering was first claimed 
to have been observed in the difference of total 
$\pi ^{\pm} d$ cross sections measured at PSI 
~\cite{pe78}, and ascribed to the $\Delta _{33}$-mass 
splitting.  Essentially, the total cross section is 
mainly determined by the forward scattering amplitude, 
which at small angles can be approximated well by 
single scattering.  In this approach, the different 
charge states of the $\Delta _{33}$ are excited in 
$\pi ^+$ and $\pi ^-$ scattering on the deuteron, and 
the result is the small observed CSV effect.  This 
has been discussed widely (see {\it {e.g.}}, the book 
by Ericson and Weise \cite{EW}).  The situation with 
the observation of CSV effects in the $\pi ^{\pm} d$ 
differential cross sections is less clear.  The first 
systematic study of the CSV effect in the differential 
$\pi ^{\pm} d$ cross sections was done at LAMPF~\cite{ma82}.  
There have been several subsequent measurements at 
LAMPF and TRIUMF for both $\pi ^+ d$ and $\pi ^- d$ 
(see Ref.~\cite{baru} for details).  The experimental 
data weakly suggest a small effect in the asymmetry 
$A_{\pi}$ for the deuteron.  For example, in 
Ref.~\cite{ma82}, an asymmetry $A_{\pi} \simeq 2\%$ 
at 143~MeV near $90^{\circ}$ in the CM frame is 
reported.  However, Smith \textit{et al.} \cite{sm88}, 
in an independent measurement done at TRIUMF, reported 
asymmetries of $-$1.5 $\pm$ 0.6\% at back angles and 
various energies.  Thus, the magnitude of CSV is at 
most 1$-$2\%, the sign is uncertain, and the experimental 
uncertainties are only slightly less than the $A_{\pi}$ 
values themselves.

At about the same time, however, measurements of 
the ratios (\ref{1}) for $^3H$ and $^3He$ at LAMPF 
obtained significantly larger effects.  The first 
evidence for a sizable CSV in the differential 
${\pi ^{\pm}}{^3H/^3He}$ cross sections
\footnote{To eliminate some of systematic 
          uncertainties, normalized yields were used 
          for the experimental determination of the 
          ratios (\ref{1}).}
below and at the $\Delta _{33}$ resonance was seen 
for the range of CM scattering angles between 
$45^{\circ}$ and $95^{\circ}$ \cite{ne90}.  The 
effect seems to peak near $80^{\circ}$ in the CM 
frame ({\it {e.g.}}, $r_2 \simeq R \simeq 1.2$ at 
180~MeV \cite{ne90}).  The experiment was repeated 
with better statistics and systematics for 
approximately the same range of scattering angles 
at energies spanning the $\Delta _{33}$ \cite{pi91} 
and beyond \cite{dh96}.  These measurements also were 
extended to backward angles from $120^{\circ}$ to 
$170^{\circ}$ \cite{ma93} and are reported in the 
previous paper \cite{ma00}.  The experimental data 
for all three ratios (\ref{1}) for incident pion 
energies between 142 and 256~MeV are shown in 
Fig.~\ref{f1}.  The agreement between the four data 
sets, on the whole, is very good.  The bump observed 
at $\sim 80^{\circ}$, corresponding to the minimum 
in the non-spin-flip amplitude, is obvious in the 
ratios $r_2$ and $R$ for 142 and 180~MeV (below and 
on the $\Delta _{33}$ resonance).  Thus, CSV effects 
in $\pi ^3H$ and $\pi ^3He$ scattering are large and 
statistically significant.  The main goal of the 
present work is to provide a theoretical basis for 
these large effects in the three-nucleon system.

\section{Amplitudes of Pion Elastic 
Scattering from $^3H$ and $^3H{\rm e}$}
\label{sec:theor1}

We formulate the pion-nuclear amplitude in the 
range of the $\Delta _{33}$ resonance as a 
combination of a single- and double-scattering 
of pions from the nucleons in the nucleus.  For the 
$A = 3$ nuclei, the appropriate diagrams are shown 
in Fig.~\ref{f2}.  The elementary $\pi N$ amplitude 
$\hat f_{\pi N}$ is taken as the $P_{33}$ partial 
wave, as if $\pi N$ scattering takes place entirely
through the $\Delta _{33}$ resonance:
\begin{eqnarray}
\hat f_{\pi N} = f_{P_{33}} \cdot\hat S \cdot\hat T , 
\label{2}\end{eqnarray}
where
\begin{eqnarray}
f_{P_{33}} = \frac{1}{2 i k_{cm}} ~[e^{2i \delta_{33} 
(k)} - 1],
\nonumber\end{eqnarray}
and $\hat S$ and $\hat T$ are the spin and isospin 
projection operators for the $\pi N$ system for 
total spin $3/2$ and isospin $3/2$:
\begin{eqnarray}
\hat S = 2 (\hat{\vec{k}}_1\cdot\hat{\vec{k}}_2) 
       + i~\vec{\sigma}\cdot [ \hat{\vec{k}}_1
       \times\hat{\vec{k}}_2]
~{\rm and}~
\hat T = \frac{1}{3} (2 + \vec t \cdot\vec{\tau}).
\label{3}\end{eqnarray}
Here, $\vec t$ and $\vec{\tau}/2$ are the isospin 
operators of the pion and nucleon, respectively, 
$\vec{\sigma}$ and $\vec{\tau}$ are Pauli matrices, 
$\hat{\vec{k}}_1$ and $\hat{\vec{k}}_2$ are the 
unit vectors in the direction of the incoming and 
outcoming pions in CM frame, respectively, and 
$\vec{k}_{cm}$ is the pion momentum in the CM 
frame.  Everywhere below, we use the following 
notation:
\begin{eqnarray}
\hat S = a + \hat b,~~
     a = 2~(\hat{\vec{k}}_1\cdot\hat
{\vec{k}}_2),~~
\hat b = ( \vec{\sigma}\cdot\vec b),
~{\rm and}~
\vec b = i~[\hat{\vec{k}}_1\times\hat
{\vec{k}}_2].
\label{4}\end{eqnarray}
The WF of the $^3H$ and $^3He$ can be written as
\begin{eqnarray}
\Psi = \psi (\vec{r}_1, \vec{r}_2, 
\vec{r}_3) 
\sum _{i = 1}^{3}X_i \cdot Y_i,
\label{5}\end{eqnarray}
where $X_i$ and $Y_i$ correspond to the spin and 
isospin parts of the WF, respectively:
\begin{eqnarray}
X_i = \frac{1}{\sqrt{2}} (\chi _i ^+ \chi)~
                         (\chi _j ^+ \sigma _2 
                          \chi _k ^*)
~{\rm and}~
Y_i = \frac{1}{\sqrt{6}} (\eta _i ^+ \vec{\tau} 
\eta)~
                         (\eta _j ^+ \vec{\tau} 
\tau _2 
                          \eta _k ^*),
\label{6}\end{eqnarray}
where $\chi$ and $\eta$ are the spinor and 
isospinor of the nucleus, respectively, and $\chi 
_i$, $\chi _j$, and $\chi _k$ ($\eta _i$, $\eta 
_j$, $\eta _k$) are the spinors (isospinors) of 
the nucleons of the nucleus ($i, j, ~{\rm and}~ 
k$ are cyclic).  Equations (\ref{6}) represent 
states of pairs of nucleons ($jk$) such that their 
total spin $S = 0$ and their isospin $T = 1$.  The 
representation (\ref{5}) for the WFs of $^3H$ and 
$^3He$ can be generalized for more complicated 
wave configurations of nuclei, as is discussed in 
Ref.~\cite{bill}.

The coordinate part of the WF is taken into 
account in a symmetric form corresponding to a 
simple $S$-shell model.  In the following 
calculations, we use two different forms of the 
radial WF: \\

i) The simple Gaussian form:
\begin{eqnarray}
\psi (\vec{r}_1, \vec{r}_2, \vec{r}_3) \sim exp 
[- \frac{1}{2 b^2}\sum _{i = 1}^{3}(\vec{r}_i - 
\vec{R}_0)^2],
\label{7}\end{eqnarray}
where $b = 1.65~fm$ and $\vec{R}_0 = \frac{1}{3} 
( \vec{r}_1 + \vec{r}_2 + \vec{r}_3)$, taken 
from Kamalov \textit{et al.} \cite{kam}.  The 
slope $b = 1.65~fm$ was chosen in Ref.~\cite{kam} 
by a best fit to the experimental data for the 
$^3H$ charge form factor below momentum transfer 
$Q = 400~MeV/c$ \cite{juster}.  This form of the 
WF fails to reproduce the minimum of the charge 
form factor at $Q \approx 710~MeV/c$; however, the 
elastic ${\pi ^{\pm}}{^3He}$ differential cross 
sections are reproduced well at $T_{\pi}$ = 100
and 200~MeV, and at backward scattering angles, 
the results of the calculations with WF~(\ref{7}) 
tend toward the suppression seen in the 
experimental cross sections \cite{kam}. \\

ii) The two-component Gaussian parametrization:
\begin{eqnarray}
\psi (\vec{r}_1, \vec{r}_2, \vec{r}_3) = N 
\sum _{m = 1}^{2}D_m~\exp[- \frac{\alpha _m}{2}
\sum _{i = 1}^{3}(\vec{r}_i - \vec{R}_0)^2],
\label{8}\end{eqnarray}
where $D_1 = 1$, $D_2 = -1.9$, $\alpha _1 = 0.70
~fm^{-2}$, $\alpha _2 = 2.24~fm^{-2}$, and N is a 
normalization constant given in Appendix~\ref{sec:CFF}.  
This WF was successfully used by Foursat \textit{et 
al.} \cite{foursat} for the description of the 
differential cross sections for the reaction $^4He(p, 
d)^3He$ at 770~MeV for a wide range of scattering 
angles.  The WF~(\ref{8}) reproduces the minimum of 
the $^3He$ charge form factor at $Q = 670~MeV/c$, 
but shows an enhancement at momentum transfer $Q 
\approx 300 - 400~MeV/c$. \\

\subsection{Single-Scattering Approximation}
\label{sec:SSA}

The diagram in Fig.~\ref{f2}a corresponds to the 
single-scattering approximation for the elastic 
pion-nuclear scattering amplitude.  To calculate 
this amplitude, we need to compute the matrix 
element for the operator (\ref{2}) between initial- 
and final-state wave functions.  We neglect both 
the Fermi motion of the nucleon inside the nucleus 
and the off-shell corrections to the $\pi N$ 
amplitudes in expression (\ref{2}).  (We discuss 
the accuracy of both of these approximations in 
Sec.~\ref{sec:ADC}.)  Taking nuclear WFs in the 
form (\ref{5}), we exclude spin and isospin 
variables of the nucleons.  As in Ref.~\cite{bill}, 
the single-scattering amplitude $\hat F_1$ is:
\begin{eqnarray}
\hat F_1 = F ( \vec\Delta )~f_{P_{33}}~\hat
\Lambda _1,
\label{9}\end{eqnarray}
where $F( \vec\Delta )$ is a nuclear form factor 
defined by Eq.~(\ref{A4}) and $\vec\Delta$ is the 
three-momentum transfer.  The operator $\hat\Lambda 
_1$ acts on the nuclear spin and isospin variables 
and is expressed as
\begin{eqnarray}
\hat\Lambda _1 = \frac{1}{3}[(6 + \vec t\cdot\vec
{\tau})~a + (2 - \vec t\cdot\vec{\tau})\hat b],
\label{10}\end{eqnarray}
where $a$ and $\hat b$ have been defined in 
Eq.~(\ref{4}).  Calculating the matrix element 
from the operator $\hat\Lambda _1$ on isotopic 
variables, we get
\begin{eqnarray}
\hat{\tilde{\Lambda}}_1 = \cases{
\frac{1}{3} (7~a + \hat b) ~~\phantom{xx}
{\rm for}~{\pi ^+}{^3He}~{\rm and}~{\pi ^-}
{^3H}~
{\rm scattering}, \cr
\frac{1}{3} (5~a + 3~ \hat b)~~~{\rm for}~
{\pi ^+}{^3H}~{\rm and}~{\pi ^-}{^3He}~
{\rm scattering}.}
\label{11}\end{eqnarray}
In terms of $\hat{\tilde{\Lambda}}_1$, the 
expression for the differential cross section 
with unpolarized particles yields:
\begin{eqnarray}
\frac {d \sigma}{d \Omega} = F^2 (\vec{\Delta})~
|f_{P_{33}} (k)|^2~\frac{1}{2}~Tr \{ 
\hat{\tilde{\Lambda}}^+_1 
\hat{\tilde{\Lambda}}_1 \},
\label{12}\end{eqnarray}
where
\begin{eqnarray}
\frac{1}{2} Tr \{\hat{\tilde{\Lambda}}^+_1 
\hat{\tilde{\Lambda}}_1 \} = \cases{
\frac{1}{9} (1 + 195~z^2)~~\phantom{xx}
{\rm for}~{\pi ^+}{^3He}~{\rm and}~{\pi ^-}
{^3H}~{\rm scattering}, \cr
\frac{1}{9} (9 + 91~z^2)~~\phantom{xxx}
{\rm for}~{\pi ^+}{^3H}~{\rm and}~{\pi ^-}
{^3He}~{\rm scattering},}
\label{13}\end{eqnarray}
where $z = ( {\hat{\vec{k}}}_1\cdot{\hat{\vec{k}}}
_2) = {\cos\theta}$.  In additional to a lower
form factor $F(\vec{\Delta})$, the angular 
dependence of $d \sigma /d \Omega$ is determined 
by the factors of Eq.~(\ref{13}).  Expressions 
(\ref{12}) and (\ref{13}) show that $d \sigma 
/d \Omega$ is suppressed at $z = 0$ ($\theta 
= 90^{\circ}$), where only the spin-flip $\pi 
N$ amplitude contributes.  Thus, for the $A = 
3$ case, there is a more significant spin-flip 
suppression than for the deuteron case, where 
(see, for example, Ref.~\cite{baru})
\begin{eqnarray}
\frac {d \sigma _{\pi d}}{d \Omega} \sim
(1 + 5~z^2)
\nonumber\end{eqnarray}
and the minimum in the cross section is much 
weaker.  We note that this kind of suppression 
of the spin-flip $\pi N$ amplitude in the 
single-scattering term for the $^3H/^3He$ case 
was pointed out in Ref.~\cite{ne90}, following 
Ref.~\cite{land}.  

In Figs.~\ref{f3} $-$ \ref{f6}, the single-scattering 
contributions to the differential cross sections 
for incident pion kinetic energies $T_{\pi}$ = 142, 
180, 220, and 256~MeV, for both versions of the 
radial part of the WFs~(\ref{7}) and (\ref{8}), 
are shown by the dashed curves.

\subsection{Double-Scattering Approximation}
\label{sec:DSA}

\begin{figure}[ht]
\centerline{
\psfig{file=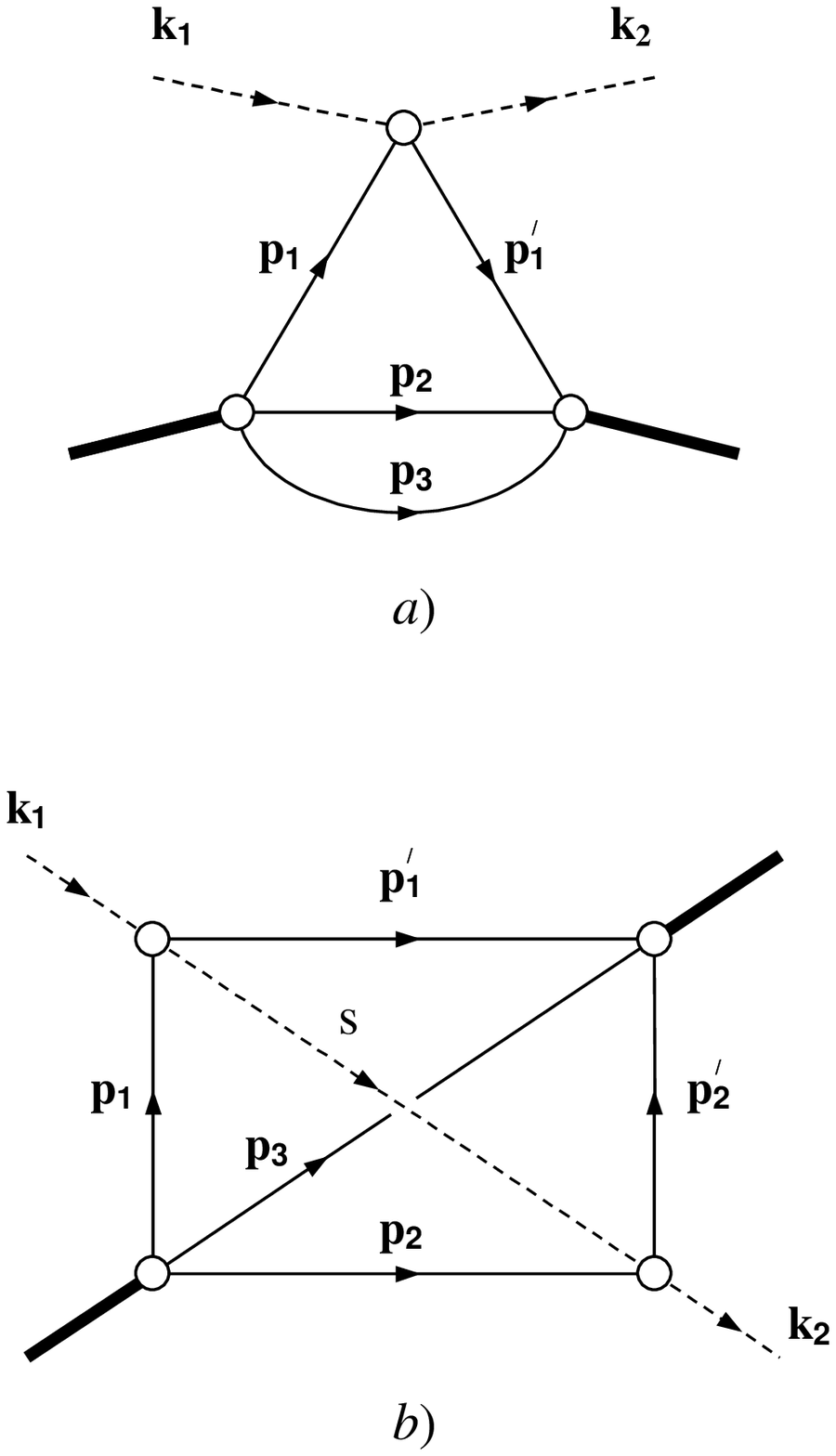,width=5in,clip=,silent=,angle=0}}
\vspace{3mm}
\caption[fig2]{\label{f2}
 (a) Single- and (b) double-scattering diagrams 
 used in the present calculations for ${\pi 
 ^{\pm}}{^3H/^3He}$ elastic scattering.}
\end{figure}

Let us consider the double-scattering amplitude 
$\hat F_2$, as in Ref.~\cite{bill}, corresponding 
to the diagram shown in Fig.~\ref{f2}b.  
(Calculations of double-spin-flip amplitudes were 
performed recently \cite{CG}.)  In the same 
approximation as was used while calculating the 
single-scattering term,
\begin{eqnarray}
\hat F_2 = 4 \pi\frac{9}{2}~f_{P_{33}}^2~\int
\frac{d^3 \vec q}{(2 \pi)^3}~\frac{d^3 \vec Q}
{(2 \pi)^3}~
\frac{d^3 \vec Q^{\prime}}{(2 \pi)^3}~\varphi 
(\vec{q^{\prime}}, \vec{Q}^{\prime})~
\varphi (\vec q, \vec Q)~G_{\pi}~
\hat{\Lambda}_2,
\label{14}\end{eqnarray}
where the WF in the momentum space operator 
$\varphi (\vec q, \vec Q)$ is defined by 
Eq.~(\ref{A2}).  The momenta $\vec q$, $\vec Q$, 
$\vec{Q^{\prime}}$, and $\vec{q^{\prime}}$ 
relate to the momenta $\vec p_i$ and $\vec 
{p^{\prime}}_i$, shown in Fig.~\ref{f2}b, via 
$\vec q = ( \vec p_2 - \vec p_3)/2$, $\vec Q 
= (\vec p_2 + \vec p_3 - \vec p_1)/3$, 
$\vec{Q^{\prime}} = (\vec{p^{\prime}}_2 + 
\vec p_3 - \vec{p ^{\prime}}_1)/3$, and 
$\vec{q^{\prime}} = \vec q - \frac{1}{2}\vec 
Q + \frac{1}{2}\vec{Q ^{\prime}} + \frac{1}
{3}\vec{\Delta}$.  In Eq.~(\ref{14}), $G_{\pi}$ 
is the Green's function for the intermediate 
state, which, neglecting kinetic energy of the 
intermediate nucleons, has the form
\begin{eqnarray}
G_{\pi} = (k_1 ^2 - {\vec{s}~}^2 + i 0)
^{-1},
\label{15}\end{eqnarray}
where $ \vec s = \vec k_1 - \vec Q + \vec 
Q^{\prime} - \frac{1}{3}\vec{\Delta}$.  By 
analogy with Eq.~(\ref{4}), we introduce $a_i$ 
and $\hat b_i$, where $i = 1, 2$, for the first 
and second nucleons:
\begin{eqnarray}
     a_{1,2} = 2~(\hat{\vec{k}}_{1,2}{\cdot}
\hat{\vec{s}} )~{\rm and}~
\hat b_{1,2} =   (\vec{\sigma}{\cdot}{\vec{b}}
_{1,2} ),
\label{16}\end{eqnarray}
where $\vec b_1 = i~[\hat{\vec{k}}_1{\times}
\hat{\vec{s}}]$, $\vec b_2 = i~[\hat{\vec{s}}
{\times}\hat{\vec{k}}_2]$, and $\hat{\vec s} 
= {\vec s}/|{\vec s}|$.  The operator $\hat
\Lambda _2$ then can be written as
\begin{eqnarray}
\hat\Lambda _2 = \frac{4}{9} (5 + \vec t
\cdot\vec{\tau}) a_1 a_2 + \frac{4}{9} (a_1 
\hat b_2 + a_2 \hat b_1) - \frac{1}{9} (6 + 
3~\vec t\cdot\vec{\tau}) \hat b_1 
\hat b_2 - \frac{1}{9} (6 + 5~\vec t\cdot
\vec{\tau}) \hat b_2 \hat b_1.
\label{17}\end{eqnarray}
The operator $\hat\Lambda _2$ depends explicitly 
on the spin and isospin variables of the nuclei 
and the pion.  $\hat\Lambda _2$ also depends on 
the vector $\hat{\vec s}$, which is an integrative 
variable on the right-hand side of Eq.~(\ref{14}).  
We extract $\hat{\vec s}$ by introducing the
operators $\hat\Lambda _{2,ij}$:
\begin{eqnarray}
\hat\Lambda _2 = \sum _{i,j}\Lambda _{2,ij} 
\hat s_i \hat s_j.
\label{18}\end{eqnarray}
Thus, the operator $\hat F_2$ can be expressed in 
the form 
\begin{eqnarray}
\hat F_2 = \sum _{i,j} f_{P_{33}}^2~\Lambda 
_{2,ij}~I_{ij},
\label{19}\end{eqnarray}
where the tensor $I_{ij}$ is expressed in the 
form
\begin{eqnarray}
I_{ij} = J_1 \hat{\kappa}_i \hat{\kappa}_j + 
J_2 \delta _{ij},
\label{20}\end{eqnarray}
such that $\vec\kappa = (\vec{k}_1 + \vec{k}
_2)/2$ and $\hat{\vec{\kappa}} = \vec{\kappa}
/| \vec{\kappa}|$.  Here, the quantities $J_1$ 
and $J_2$ are complex functions which depend on 
the momentum of the incoming pion $k_1$ and 
the momentum transfer $\Delta$ (or on the 
scattering angle $\theta$).  They also depend 
on the WFs of the nuclei, which are given in 
Appendix~\ref{sec:tensor}.  Using Eq.~(\ref{20}) 
for the tensors $I_{ij}$, the amplitude $\hat 
F_2$ becomes
\begin{eqnarray}
\hat F_2 = A_2 + i~B_2~(\vec\sigma\cdot 
[\hat{\vec{k}}_1{\times}\hat{\vec{k}}_2]).
\label{21}\end{eqnarray}
We divide the contributions to $A_2$ and $B_2$ 
into double elastic scattering 
\begin{eqnarray}
\pi ^- \to \pi ^- \to \pi ^-
\nonumber\end{eqnarray}
(we will define this contribution via index 
``$ee$") and double charge exchange 
\begin{eqnarray}
\pi ^- \to \pi ^0 \to \pi ^-
\nonumber\end{eqnarray}
(we will denote this contribution via index 
``$cc$") in the functions $A_2$ and $B_2$:
\begin{eqnarray}
A_2 = A_2^{ee} + A_2^{cc}~{\rm and}~
B_2 = B_2^{ee} + B_2^{cc}.
\label{22}\end{eqnarray}
Then, we can present expressions for $A_2$ and 
$B_2$ in terms of the integrals $J_1$ and $J_2$:
\begin{eqnarray}
A_2^{cc} = - \frac{1}{9}f_{P_{33}}^2 [(3 + 5 z)
J_1 + 12 z J_2],~~
B_2^{ee} = \frac{4}{3}f_{P_{33}}^2 (J_1 + 2 J_2),
\phantom{xxxxxxxxxxxxxxxxxx}\cr
A_2^{ee} = \frac{1}{3}f_{P_{33}}^2 [(23 + 17 z) 
J_1 + 
28 z J_2],~~B_2^{cc} = - \frac{2}{9}f_{P_{33}}^2 
(2 J_1 + 3 J_2)~{\rm for}~
{\rm {\pi ^+}{^3He},~{\pi ^-}{^3H}}, \cr
A_2^{ee} = \frac{1}{9}f_{P_{33}}^2 [(29 + 27 z) 
J_1 +
52 z J_2],~~B_2^{cc} = - \frac{2}{9}f_{P_{33}}^2
(2 J_1 + 5 J_2)~{\rm for}~
{\rm {\pi ^-}{^3He},~{\pi ^+}{^3H}}. 
\label{23}\end{eqnarray}

The resulting double-scattering contributions to 
the differential cross sections are shown by the
dotted curves in Figs.~\ref{f3} $-$ \ref{f6}.

\subsection{Nonresonant Contributions}
\label{sec:NRC}

Although, the $\pi N$ $P$-wave amplitude is 
dominant for the multiple scattering of pions 
on the nucleons of nucleus in the energy range 
under consideration, the contribution of small 
nonresonant waves still can play an important 
role near $\theta \sim 90^{\circ}$ (the sharp 
minimum for single scattering).  Thus, we limit 
ourselves to single scattering for nonresonant 
waves.  In this limit, the $\pi N$ amplitude 
(\ref{2}) becomes
\begin{eqnarray}
\hat f_{\pi N} = \sum_{j}f_j \cdot\hat S_j 
\cdot\hat T_j,
\label{24}\end{eqnarray}
where
\begin{eqnarray}
f_j = \frac{1}{2 i k_{cm}}[e^{2 i \delta _j 
(k)} - 1]
\nonumber\end{eqnarray}
and $\hat S_j$ and $\hat T_j$ are the spin and 
isospin projection operators for the state $j$ 
of the $\pi N$ system.  Limiting ourselves to 
$S$- and $P$-waves, the nonresonant $\pi N$ 
$\delta _j (k)$ phases can be taken from a 
recent GW $\pi N$ partial-wave analysis 
\cite{gwpwa}.  
Therefore, we take into account two $S$- and 
four $P$-waves in our calculations.  The 
projection operators are
\begin{eqnarray}
\hat S_j = \hat 1
\phantom{xxxxxxxxxxxxxxx}(S_{11}, S_{31}),
\phantom{xxxxxxx}\cr
\hat S_j = z - i (\vec{\sigma}\cdot\vec n)
\phantom{xxxxxxx}(P_{11}, P_{31}),
\phantom{xxxxxxx}\cr
\hat S_j = 2 z + i (\vec{\sigma}\cdot\vec n)
\phantom{xxxxxx}(P_{13}, P_{33}), 
\phantom{xxxxxxx}\cr
\hat T_j = \frac{1}{3}(1 - 
\vec t\cdot\vec{\tau})
\phantom{xxxxxxx}(S_{11}, P_{11}, P_{13}), 
~{\rm and}~ \cr
\hat T_j = \frac{1}{3}(2 + 
\vec t\cdot\vec{\tau})
\phantom{xxxxxxx}(S_{31}, P_{31}, P_{33}),
\phantom{xxxx}
\label{25}\end{eqnarray}
where $\vec n = [\hat{\vec k_1}\times\hat{\vec 
k_2}]$.  We note that the procedure of taking 
into account nonresonant waves in the 
single-scattering approach is analogous to the 
resonant contribution taken into account and 
discussed in Sec.~\ref{sec:SSA}.  The final 
expressions for the non-spin-flip and spin-flip 
nonresonant amplitudes are
\begin{eqnarray}
A_{non} = \left[ \frac{2}{3} f_{S_{11}} + 
\frac{7}{3} f_{S_{31}} +
\left(\frac{2}{3}f_{P_{11}} + \frac{7}{3}
f_{P_{31}} + \frac{4}{3}f_{P_{13}}\right)
 z \right] F(\Delta)\, ~{\rm and}~
\cr
B_{non} = \frac{1}{3}\,(2 f_{P_{13}} - f_{P
_{31}} - 2 f_{P_{11}})\, F(\Delta)\,
\phantom{x} ~{\rm for}~
\pi^+\,^3{\rm He}\, ~{\rm and}~ \pi^-\,^3{\rm H},
~{\rm and}~\cr
A_{non} = \left[ \frac{4}{3} f_{S_{11}} + 
\frac{5}{3} f_{S_{31}} +
\left(\frac{4}{3}f_{P_{11}} + 
\frac{5}{3}f_{P_{31}} + \frac{8}{3}f_{P_{13}}
\right)
 z \right] F(\Delta)
\phantom{x}
~{\rm and}~\cr
B_{non} = - f_{P_{31}}\, F(\Delta)\,
\phantom{xxxxxxxxxxxxxxxxxxx}
~{\rm for}~
\pi^-\,^3{\rm He}\, ~{\rm and}~ \pi^+\,^3{\rm H}.
\label{26}\end{eqnarray}
The nonresonant contribution for the total 
amplitude $\hat F = \hat F_1 + \hat F_2$ (see 
Eq.~(\ref{27}) below) is expressed by the 
substitutions
\begin{eqnarray}
A\to A + A_{non} ~{\rm and}~ B\to B + 
B_{non}.
\nonumber\end{eqnarray}
The nonresonant amplitudes are taken into account 
in the calculations of the ratios (\ref{1}) 
presented in Sec.~\ref{sec:theor3}.

\subsection{Total Amplitude and Differential 
Cross Section}
\label{sec:DSG}

The expression for the sum of the single- and 
double-scattering amplitudes can be expressed 
in a form similar to Eq.~(\ref{21}):
\begin{eqnarray}
\hat F = A + i~B~(\vec\sigma\cdot [\hat
{\vec k}_1\times\hat{\vec k}_2]),
\label{27}\end{eqnarray}
where the functions $A$ and $B$ represent the 
contributions of single and double scattering, 
{\it {e.g.}}, $A = A_1 + A_2$ and $B = B_1 + 
B_2$.  The amplitudes $A_1$ and $B_1$ are 
determined by Eqs.~(\ref{9}) $-$ (\ref{11}) and 
(\ref{26}); $A_2$ and $B_2$ by Eqs.~(\ref{22}) 
$-$ (\ref{23}).  In terms of the functions $A$ 
and $B$, the differential cross section in 
unpolarized case has the form
\begin{eqnarray}
\frac {d \sigma}{d \Omega} = \frac{1}{2}Tr 
\{\hat F^+ \hat F \} = |A|^2 + |B|^2 \sin 
^2 \theta.
\label{28}\end{eqnarray}

The combined single- and double-scattering
contributions, with interference taken into 
account, are depicted by the solid curves in 
Figs.~\ref{f3} $-$ \ref{f6}.  It can be seen that 
the model approach we use qualitatively agrees 
with the data.  For forward scattering, $\theta
\sim 30 - 60^{\circ}$, WF~(\ref{7}) reproduces 
the cross sections systematically better than 
WF~(\ref{8}).  At larger scattering angles, the 
results for WF~(\ref{7}) (WF~(\ref{8})) lie 
below (above) the experimental data, and the 
cross-section minimum is shifted to a smaller 
angle ($\theta\sim 80^{\circ}$), in agreement 
with the experimental data.  The key point to be 
made, however, is that the gentle maximum at 
$\theta\sim 110 - 120^{\circ}$ arises from the 
interference between single and double scattering.  
It is also seen that the moderation of the rise 
in the cross sections for backward scattering 
reflects the contribution of double scattering, 
which is most pronounced near $\theta\sim 180^{\circ}$.

We point out that the description of hadron-nuclear 
scattering in the backward hemisphere is a very 
complicated multibody problem that requires detailed 
information about the wave function of the nucleus 
and the reaction mechanism.  For example, the 
5-component WF used in Ref.~\cite{kam} to describe 
the ${\pi ^{\pm}}{^3He}$ elastic cross sections at 
$T_{\pi}$ = 200~MeV is 
superior to WF~(\ref{7}).  However, the goal of our 
study is the description and understanding of CSV 
effects, and these effects do not depend strongly
on the details of the WF at small distances.  For 
this reason, we prefer to use the simpler $S$-shell 
versions of the nuclear WFs~(\ref{7}) and (\ref{8}).

\section{Charge-Symmetry Violation Effects}
\label{sec:theor2}

There are three principal sources of the violation of 
charge symmetry for $\pi ^{\pm}$ scattering from light 
nuclei in the $\Delta _{33}$ region:

\begin{enumerate}
\item
the Coulomb interaction between the charged pions and 
the nuclei $-$ the external Coulomb  effect,
\item
the mass splitting of the different charge states of 
the $\Delta _{33}$-isobar, and
\item
the difference between the WFs of $^3H$ and $^3He$ due 
to the additional Coulomb repulsion between the two 
protons in the $^3He$ nucleus $-$ the internal Coulomb 
effect.
\end{enumerate}

We now discuss how we take these effects into account 
in our calculation of the elastic scattering of 
charged pions from the $A = 3$ nuclei.

\subsection{External Coulomb Effect}
\label{sec:ECI}  

As was shown in Sec.~\ref{sec:expt}, experimental 
data for the ratios (\ref{1}) \cite{ne90,pi91,dh96,ma00} 
were taken outside the Coulomb cone, $\theta\geq 
30^{\circ}$.  In this angular range, the Coulomb 
amplitude is a smooth function of the scattering 
angle $\theta$.  Here, we take into account the 
Coulomb interaction in a nonrelativistic approach, 
neglecting the interaction between the photon and 
the magnetic moment of the nucleus.  Thus, the 
Coulomb amplitude of the pion-nucleus interaction 
in terms of the Coulomb phase may be written as
\begin{eqnarray}
A_C = - \frac{Z_{\pi} Z_A e^2}{2 k_{cm}^2 \sin ^2 
\frac{\theta}{2}} \frac{\omega m_A}{m_A + 
\omega}\exp[- \frac{2 i Z_{\pi} Z_A e^2}{k_{cm}}
\frac{\omega m_A}{m_A + \omega}\ln(\sin\frac
{\theta}{2})]~F_{\pi} (\vec{\Delta})~F(\vec{\Delta}),
\label{29}\end{eqnarray}
where $e^2 \simeq 1/137$, $Z_{\pi}$ and $Z_A$ are 
the charges of the pion and the nucleus, respectively, 
$\omega$ is the pion CM energy, $m_A$ is the mass of 
the nucleus, F($\vec{\Delta}$) is the form factor of 
the nucleus defined by the Eq.~(\ref{A4}), and 
$F_{\pi}(\vec{\Delta})$ is the pion charge form factor 
that is used in the standard parametrization of 
Ref.~\cite{pionff}.  In calculating the ratios 
(\ref{1}), we use the amplitude $A_C$ of 
Eq.~(\ref{29}) in combination with the non-spin-flip 
amplitude $A$ of the strong interaction of 
Eq.~(\ref{27}) by the substitution $A\to A + A_C$.

\subsection{$\Delta _{33} (1232)$-Mass
Splitting}
\label{sec:DMS}

The influence of the $\Delta _{33}$-mass splitting 
on the differential cross section for $\pi d$ 
elastic scattering was discussed in Ref.~\cite{ma82}, 
where the single-scattering approximation with 
allowance for the different charge states of the 
$\Delta _{33}$(1232) was used.  In this 
approximation, the CSV effect proves to be independent 
of the scattering angle, with a value proportional 
to $\delta m_{\Delta} / \Gamma _{\Delta}$.  Nearly 
the same approach was used for the $^3H/ ^3He$ case 
in Ref.~\cite{ne90}.

We denote the different charge states of the $\Delta 
_{33}$ via the index $i = 1,~~2,~~3,$ and $4$ for 
the $\Delta ^{++},~~\Delta ^{+},~~\Delta ^{0}$, and
$\Delta ^{-}$, respectively.  The mass $w_i~(i = 
1,~2,~3,~4)$, corresponding to the isobar $i$, is 
calculated according to a formula from Ref.~\cite{EW} 
(page~109, Eq.~(4.18)), following Ref.~\cite{WT}: 
\begin{eqnarray}
w_i = a - b~I_i + c~I^2_i,
\label{30}\end{eqnarray}
where $I_i$ is the 3$^{rd}$ component of isospin for 
the $i^{th}$-term from the $\Delta _{33}$-multiplet.  
Using the average resonance mass value from the 
Particle Data Group \cite{PDG}, $\bar w = 1232~MeV$, 
$b = 1.38~MeV$ from Ref.~\cite{EW}, and 
\begin{eqnarray}
w_3 - w_1 = m_{\Delta ^0} - m_{\Delta ^{++}} \simeq 
2.5~MeV
\nonumber\end{eqnarray}
from Ref.~\cite{PDG}, we get
\begin{eqnarray}
a \simeq 1231.8~MeV ~{\rm and}~ c = 0.13~MeV.
\nonumber\end{eqnarray}
The scalar amplitude for $\pi N$ scattering (see 
Eq.~(\ref{2})) for each charge state $i$ is 
defined as
\begin{eqnarray}
f_{P_{33}}\to f_i = \frac{1}{2 i k_{cm}} 
[e^{2 i \delta _i(k)} - 1].
\label{31}\end{eqnarray}

The phases $\delta _i$ are defined relative to 
the resonance phase $\delta _{P_{33}}$ 
\begin{eqnarray}
\delta _i = \delta _{P_{33}} - 2 \frac{\delta w_i}
{\Gamma _{\Delta}}\sin ^2 \delta _{P_{33}},
\label{32}\end{eqnarray}
where $\delta w_i = w_i - \bar w$.  The resonance 
phase $\delta _{P_{33}}$ is taken from \cite{gwpwa}.  
In Eq.~(\ref{32}), we neglect the energy dependence 
of the width $\Gamma _{\Delta}$ because $\delta w_i/ 
\Gamma _{\Delta}\ll 1$; in our calculations we use 
$\Gamma _{\Delta} = 120~MeV$.  Using this definition 
of $f_i$ for the $\pi N$ scattering amplitudes in 
Eq.~(\ref{31}), we obtain the following expressions 
for the single pion-nucleus scattering amplitudes:
\begin{eqnarray}
A_1 = (2 f_1 + \frac{1}{3}f_2) 2 z F(\vec
\Delta) ~{\rm and}~
B_1 = \frac{1}{3}f_2 F(\vec\Delta) ~{\rm for}~
{\rm {\pi ^+}{^3He}} ~{\rm and}~
\label{33}\end{eqnarray}
\begin{eqnarray}
A_1 = (f_1 + \frac{2}{3}f_2) 2 z F(\vec
\Delta) ~{\rm and}~
B_1 = f_1 F(\vec\Delta) ~{\rm for}~
{\rm {\pi ^+}{^3H}}.
\label{34}\end{eqnarray}
Substituting $f_1\to f_4$ and $f_2\to f_3$, the 
amplitudes $A_1$ and $B_1$ at Eqs. (\ref{33}) and 
(\ref{34}) are transformed to the amplitudes for 
the isomirror reactions ${\pi ^-}{^3H}$ and ${\pi 
^-}{^3He}$.

For the double-scattering pion-nucleus amplitudes, 
we obtain
\begin{eqnarray}
A_2^{cc} = - \frac{1}{9}f_2^2 [(3 + 5 z) 
J_1 + 12 z J_2] ~{\rm and}~
B_2^{ee} = \frac{4}{3}f_1 f_2 (J_1 + 
2 J_2) ~{\rm for}~
{\rm {\pi ^+}{^3He},{\pi ^+}{^3H}}
\label{35}\end{eqnarray}
instead of Eq.~(\ref{23}).  The other amplitudes 
for ${\pi ^+}{^3He}$ and ${\pi ^+}{^3H}$ elastic 
scattering are different:
\begin{eqnarray}
A_2^{ee} = \frac{4}{3}f_1 f_2 [2 (1 + z) J_1 + 
4 z J_2] + f_1^2 [(5 + 3 z) J_1 + 4 z J_2]
\phantom{xx}
~{\rm for}~ {\pi ^+}{^3He}, 
\phantom{x}\cr
A_2^{ee} = \frac{4}{3}f_1 f_2 [2 (1 + z) J_1 + 
4 z J_2] + \frac{1}{9}f_2^2 [(5 + 3 z) J_1 + 
4 z J_2]
\phantom{x}
~{\rm for}~ {\pi ^+}{^3H}, 
\phantom{xx}\cr
B_2^{cc} = - \frac{2}{9}f_2^2 (2 J_1 + 3 J_2)
\phantom{xxxxxxxxxxxxxxxxxxxxxx}
~{\rm for}~ {\pi ^+}{^3He},
~{\rm and}~
\phantom{xx}\cr
B_2^{cc} = - \frac{2}{3}f_2^2 (2 J_1 + 5 J_2)
\phantom{xxxxxxxxxxxxxxxxxxxxxxxxxx}
~{\rm for}~ {\pi ^+}{^3H}.
\phantom{xxx}
\label{36}\end{eqnarray}
The amplitudes $A_2$ and $B_2$ of Eqs. (\ref{35}) 
and (\ref{36}) are also transformed to their 
isomirror reaction amplitudes, {\it {i.e.}},
${\pi ^+}{^3He}\to {\pi ^-}{^3H}$ and ${\pi ^+}{^3H}
\to {\pi ^-}{^3He}$, by substituting $f_1\to f_4$ and 
$f_2\to f_3$.  We do not include any difference for 
the different charge states in the nonresonant 
amplitudes $A_{non}$ and $B_{non}$.

\subsection{Internal Coulomb Effect}
\label{sec:ICI}

The difference in the structure of the WFs of $^3H$ 
and $^3He$ is related not only to the electromagnetic 
interaction, but also to the the part of the strong 
interaction which violates isospin.  In the 
strong-interaction sector, there are terms that 
violate isospin directly \cite{wein}.  Isospin 
violation inside nuclei can relate to both nucleon 
and quark degrees of freedom.  In terms of quark 
degrees of freedom, isospin violation relates to 
the mass difference of the $u$ and $d$ quarks.  At 
present, there is no quantitatively good estimate 
of isospin violation due to the strong interaction 
for the $A = 3$ nuclei.

If we assume that the strong interaction conserves 
isospin, then the main reason for the difference in 
the structure of the WFs of $^3H$ and $^3He$ is the 
additional Coulomb repulsion between the two 
protons in $^3He$, which is not present for $^3H$.

If there were no Coulomb interaction between these 
two protons ({\it {i.e.}}, if the WFs of $^3H$ and 
$^3He$ were isotopically symmetric), the neutron 
distribution for $^3He$ (the ``odd" neutron) would 
be the same as the proton distribution for $^3H$ 
(the ``odd" proton), and the proton distribution 
for $^3He$ (the ``even" protons) would be the same 
as the neutron distribution for $^3H$ (the ``even" 
neutrons).  However, the proton distributions for 
$^3H$ and $^3He$ can still be different and, as a 
consequence, so can the charge form factors of 
$^3H$ and $^3He$. 

If, however, isospin is violated for $^3H$ and 
$^3He$, the even- and odd-nucleon distributions can 
also be different.  In Ref.~\cite{GG}, the 
difference between these distributions has been 
formulated in terms of nonzero parameters $\delta 
_e$ and $\delta _o$, where
\begin{eqnarray}
\delta _e = r_e^n - r_e^p ~{\rm and}~
\delta _o = r^n_o - r^p_o.
\nonumber\end{eqnarray}
Here, $r^{n,p}_{e,o}$ is the neutron (n) or
proton (p) radius for the even (e) or odd (o) 
nucleons.  As was shown in Ref.~\cite{GG}, 
the superratio $R$ at $T_{\pi} = 180~MeV$ is 
very sensitive to $\delta _e$ and $\delta _o$; 
a reasonable description of $R$ versus 
scattering angle has been obtained for $\delta 
_e = -0.030 \pm 0.008~fm$ and $\delta _o = 
0.035 \pm 0.007~fm$.  These differences 
between $r^n_e$ and $r_e^p$ or between $r_o^n$ 
and $r_o^p$ due to isospin violation result 
in additional changes in the charge radii and 
form factors of $^3H$ and $^3He$.

We vary the parameters of the WFs~(\ref{7}) and 
(\ref{8}) to introduce such differences.  For 
WF~(\ref{7}), we follow the recipe suggested in 
Ref.~\cite{kam}: we fix the slope $b$ for $^3H$ at 
$1.65~fm$ and vary the slope for $^3He$ to 
obtain the best description of the data for the 
ratios $r_1$ and $r_2$.  An analogous procedure 
is followed for WF~(\ref{8}), using parameters 
for the WF for $^3He$ suggested in 
Ref.~\cite{foursat} (see Sec.~ \ref{sec:theor1}).  
Then, for the WF of $^3H$, we use three different 
variations of the WF: (i) a variation of the 
slope $\alpha _1$, (ii) a variation of the slope 
$\alpha _2$, and (iii) a variation of both slopes 
$\alpha _1$ and $\alpha _2$, which are proportional 
to each other: $\alpha _1\to c \alpha _1$ and 
$\alpha _2\to c \alpha _2$.

Although this variation of the parameters of 
WFs~(\ref{7}) and (\ref{8}) cannot be compared 
directly with the refined procedure used in 
Ref.~\cite{GG}, this way of taking into account the 
internal Coulomb interaction allows us to take 
the the $\Delta _{33} (1232)$-mass splitting 
into account, which was not done in Ref.~\cite{GG}.  
Therefore, the quantities $\delta _e$ and $\delta _o$ 
obtained from the experimental data can differ from 
the values obtained in Ref.~\cite{GG}. 

\section{Comparison with the Experimental Data}
\label{sec:theor3}

\subsection{Excluding the Internal Coulomb Effect}
\label{sec:EIC}

The results of our calculations for the ratios 
$r_1$, $r_2$, and the superratio $R$ which take
into account the external Coulomb interaction 
and the $\Delta _{33}$-mass splitting but exclude 
the internal Coulomb effect are shown in 
Figs.~\ref{f7} $-$ \ref{f10}, as the dotted 
curves for single scattering and the dashed 
curves for both single and double scattering.  
There are no free parameters associated with these 
results.  The main purpose of Figs.~\ref{f7} $-$ 
\ref{f10} (as for Figs.~\ref{f3} $-$ \ref{f6}) is 
to show that the inclusion of double scattering 
is essential to be able to follow the trend of 
the data in the non-spin-flip-dip region.  As is 
seen in these figures, there is qualitative 
agreement between the results of our calculations 
and the data for $T_{\pi}$ = 180, 220, and 256~MeV.  
Also, in contrast with the case for the 
differential cross section (see Figs.~\ref{f3} $-$ 
\ref{f6}), there is little sensitivity to the WF 
here.  For $T_{\pi}$ = 180~MeV, the ratio $r_1$ 
is reproduced very well for both WFs~(\ref{7}) and 
(\ref{8}), but the peaks near $\theta = 80^{\circ}$ 
in the ratio $r_2$ and superratio $R$ are 
reproduced better by WF~(\ref{7}).  At the same 
time, the description of $r_2$ and $R$ in the 
backward direction is not good for either WF.  
Figure~\ref{f8} shows that even taking into account 
single scattering and the external Coulomb 
interaction cannot reproduce the experimental 
data there.  But overall, Figs.~\ref{f8} $-$ 
\ref{f10} show that taking into account the 
$\Delta _{33}$-mass splitting consistently with 
both the single- and double-scattering 
contributions reproduces the main structures of 
the angular distribution and that we do not 
require a detailed knowledge of the nuclear WF.

We note here that the data for $r_2$ and $R$ for 
$T_{\pi}$ = 142~MeV are not reproduced by our 
model approach.  Apparently, some other mechanism 
must play a role at this energy.  We return to 
this problem below.

\subsection{Including the Internal Coulomb Effect}
\label{sec:IIC}

We now describe the CSV effects by taking into 
account the internal Coulomb interaction as well.  
The procedure for variation of the WFs is described 
above.  For this case, we vary a single free 
parameter to obtain the best fit.  Again, we fit 
$r_1$ and $r_2$ only, since the superratio $R$ ($= 
r_1\cdot r_2$) is not an independent quantity.  
The best-fit results for this approach and for 
both WFs~(\ref{7}) and (\ref{8}) are shown in 
Figs.~\ref{f11} $-$ \ref{f14} by the solid curves.  
By comparison, the results without the internal 
Coulomb interaction (from Figs.~\ref{f7} $-$ 
\ref{f10}) are shown by the dashed 
curves.  Both free parameters $\alpha _1$ and 
$\alpha _2$ have been varied simultaneously, 
following our prescription (iii) of 
Sec.~\ref{sec:theor2}.  The results of varying 
either $\alpha _1$ or $\alpha _2$ independently 
are very similar, as listed in the Table.  In 
this Table, the best-fit results obtained by 
variations of the radii of the nuclei and 
corresponding $\chi ^2$ values are listed as well.

In Fig.~\ref{f12}, we see that our calculations 
reproduce all the ratios $r_1$, $r_2$, and $R$ for 
T$_{\pi}$ = 180~MeV rather well over the entire 
angular range.  In fact, we reproduce the 
superratio $R$ for $40^{\circ}\leq\theta\leq 
110^{\circ}$ much as Gibbs and Gibson did \cite{GG}.  
Taking into account the difference in the WFs of 
$^3H$ and $^3He$ results in a much better 
reproduction both of the quantity $R$ at 
$\theta\sim 80^{\circ}$ and the scattering at 
backward angles.  Thus, taking into account the 
internal Coulomb interaction provides a substantial 
improvement in $\chi ^2$ compared with the case 
where only the external Coulomb interaction and 
the $\Delta _{33}$-mass splitting are included.

Finally, we consider the description of the data 
for $T_{\pi}$ = 220 and 256~MeV to be 
qualitatively satisfactory, while the description 
of the data for $T_{\pi}$ = 142~MeV is not.  
Consider the situation for $T_{\pi}$ = 220~MeV, 
shown in Fig.~\ref{f13}.  Our theoretical curves 
reproduce the sharp dip-bump angular dependence 
of the ratios $r_2$ and $R$ remarkably well.  At 
$T_{\pi}$ = 256~MeV, shown in Fig.~\ref{f14}, we 
predict a more gradual dip-bump angular 
dependence, which also agrees with the experimental 
data.  At the same time, our theoretical curves do 
not reproduce the data for $T_{\pi}$ = 142~MeV, 
shown in Fig.~\ref{f11}, even though the angular 
dependence is much smoother.  We note, however, that 
the amount of experimental data for $T_{\pi}$ = 220 
and 256~MeV is significantly less than for $T_{\pi}$ 
= 142 and especially for 180~MeV.  Evaluation of the 
$T_{\pi}$ = 142~MeV data may require an additional 
mechanism to reproduce the behavior of the data.
\footnote{In the most advanced study \cite{GG}, the
          authors considered only the case for
          $T_{\pi}$ = 180~MeV.  On the other hand,
          the calculation of Ref.~\cite{kam} does not
          reproduce the CSV effect for 142~MeV either.}

\section{Additional Remarks}
\label{sec:add}

\subsection{$\Delta _{33}$-Mass Splitting and Total 
Cross Sections}
\label{sec:SGT}

\begin{figure}[ht]
\centerline{
\psfig{file=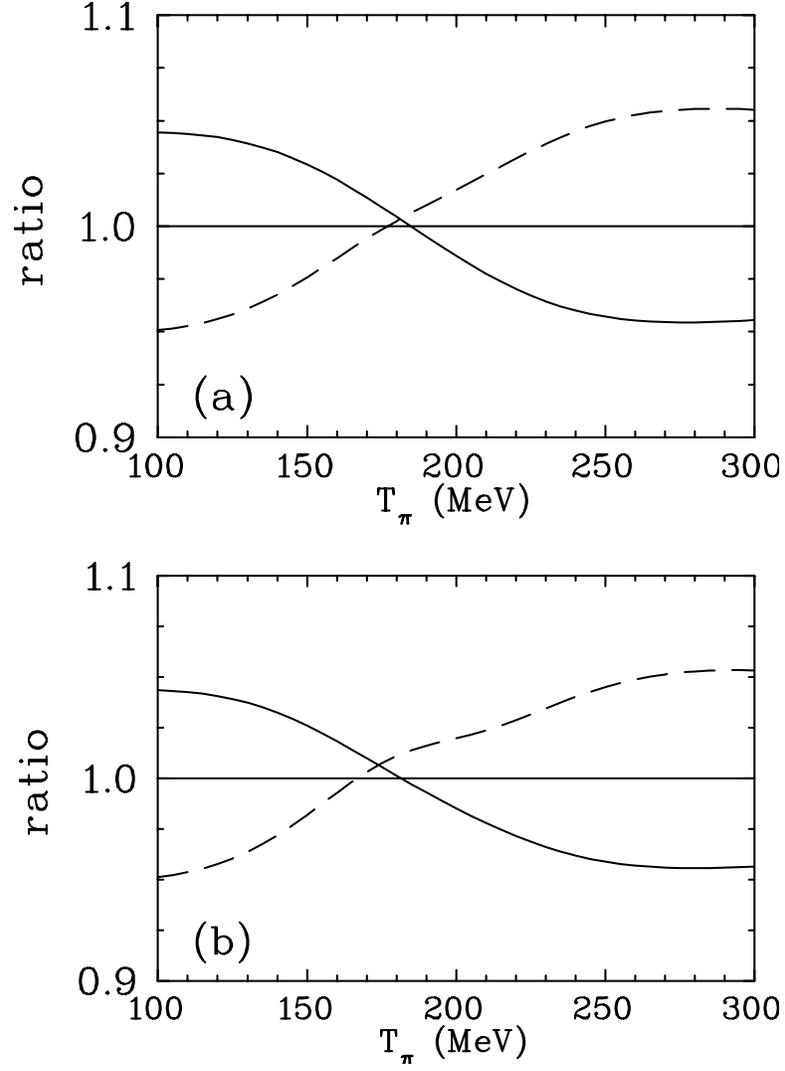,width=4in,clip=,silent=,angle=90}} 
\vspace{3mm}
\caption[fig15]{\label{f15}
 Predictions for the ratios $r_{1t}$ (solid),
 and $r_{2t}$ (dashed) for the total ${\pi
 ^{\pm}}{^3H/^3He}$ cross sections.
 Calculations were done for single and
 double scattering with the $\Delta
 _{33}$-mass splitting.  The Coulomb
 interactions are not taken into account.
 Plotted are the results for (a) WF~(\ref{7})
 and (b) WF~(\ref{8}).}
\end{figure}

We have seen that one of the principal sources of 
CSV is the $\Delta _{33}$-mass splitting.  The 
difference in the total $\pi ^{\pm} d$ cross 
sections is determined by the $\Delta _{33}$-mass 
splitting \cite{EW}.  Here, we calculate this 
effect for the ratios $r_{1t}$ and $r_{2t}$ for 
the {\it total} cross sections of ${\pi ^{\pm}}
{^3H/^3He}$ scattering, defined as
\begin{eqnarray}
r_{1t} = \frac
{\sigma ^{tot}({\pi ^+}{^3H})}
{\sigma ^{tot}({\pi ^-}{^3He})}~{\rm and}~
r_{2t} = \frac
{\sigma ^{tot}({\pi ^-}{^3H})}
{\sigma ^{tot}({\pi ^+}{^3He})}.
\nonumber
\end{eqnarray}
We predict a considerably larger CSV effect for the 
three-nucleon system than for the deuteron, as shown 
in Fig.~\ref{f15}.  Moreover, the crossover of 
$r_{1t}$ and $r_{2t}$ at the peak of the $\Delta _{33}$, 
for either of the WFs used, establishes a unique 
signature for this effect.

The high sensitivity of CSV to the $N^*$-mass splitting 
(within multiplets) allows us to suggest the use of 
these ratios as a method to determine the mass-splitting 
in heavy $N^*$s and $\Delta ^*$s.  We think that this 
method would be preferable to the traditional $\pi N$ 
partial-wave analysis for baryon spectroscopy.

\subsection{Accuracy of Calculations}
\label{sec:ADC}

In our theoretical approach to single and double 
scattering, we neglect the Fermi motion of the 
nucleons inside the nucleus.  The amplitude of the 
$\pi N$ scattering is extracted from an integral 
over incident pion energies that are on-shell.  We 
also neglect the recoil of the nucleons in the 
Green's function $G_{\pi}$ of Eq.~(\ref{14}) for the 
double-scattering amplitude.  We call this approach 
the ``fixed-centers approximation."  Then, we observe
that when only certain corrections are taken into 
account, we obtain worse agreement with the 
experimental data (see Ref.~\cite{bill} for details).

Since the fixed-centers approximation gives results 
that are close to the data, there must be 
cancellations of the main corrections to leading 
order.  These cancellations for hadron-deuteron 
scattering have been discussed in detail in previous 
work.  The cancellation of nonadiabatic corrections 
within the Glauber approach for differential cross 
sections at high energies was shown in Ref.~\cite{tyk}.  
In the $\pi d$ scattering-length calculations, the 
cancellation of off-shell and recoil corrections was 
discussed in Ref.~\cite{aek}.  The cancellation of 
nonadiabatic corrections for $\pi d$ elastic 
scattering was found in the range of the $\Delta 
_{33}$ resonance in Ref.~\cite{aek1}.  Apparently, 
the analogous cancellation of corrections holds as 
well for the $A = 3$ nuclei.  Therefore, the use of 
only a few of the corrections to the fixed-centers 
approximation can result in worse agreement with the 
data than when all the corrections are ignored.  
Inclusion of all the corrections (including the 
so-called binding corrections) is a rather 
complicated task and is not a goal of our present 
study.

For example, we limit ourselves to the consideration 
of single- and double-scattering terms, and this 
approach allows us to take into account the leading 
terms of the amplitude of the pion-nuclear 
interaction at energies and scattering angles where 
we can qualitatively reproduce the shape of the 
distribution of the differential cross sections.  
Triple-scattering contributions to the differential 
cross sections can smooth the angular shape, but 
they are more difficult to calculate and cannot be 
simplified by the transformation of the integrals 
$J_1$ and $J_2$ discussed in 
Appendix~\ref{sec:tensor}.  

\subsection{Pion Absorption}
\label{sec:PA}

\begin{figure}[ht]
\centerline{
\psfig{file=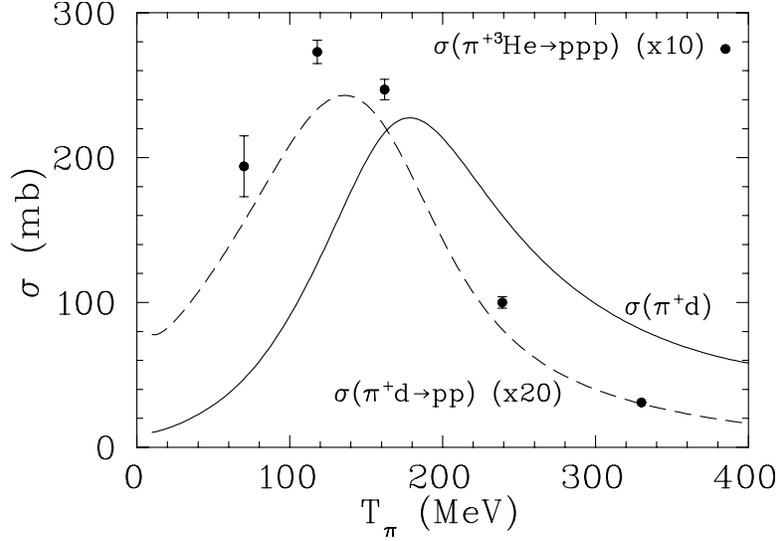,width=4in,clip=,silent=,angle=90}} 
\vspace{3mm}
\caption[fig16]{\label{f16}
 Total $\pi ^+$ cross sections.  Plotted are
 cross sections for $\pi ^+d$ (solid) and
 $\pi ^+d\to pp$ (multiplied by a factor of 20)
 (dashed) from a recent combined fit of the
 $pp$ and $\pi d$ elastic scattering with the 
 $\pi ^+d\to pp$ data \protect\cite{pdpwa}.  The
 $\pi ^3He$ absorption data (multiplied by a
 factor of 10) are from \protect\cite{psi}
 (filled circles); it can be seen that they peak
 near T$_{\pi}$ = 140~MeV.}
\end{figure}

At small scattering angles, $\theta\sim 30^{\circ} 
- 60^{\circ}$, there is reasonable agreement 
between our theoretical approach and the 
experimental data for $T_{\pi}$ = 180~MeV and above, 
but for $T_{\pi}$ = 142~MeV the experimental cross 
sections are smaller than the results of our model 
calculations for both WFs used (Figs.~\ref{f3} $-$ 
\ref{f6}).  But because for modest scattering angles 
the range of momentum transfer is small ($Q \leq 
150~MeV/c$), our use of the simple WFs~(\ref{7}) and 
(\ref{8}) is reasonable.  We therefore infer that 
the suppression of the cross section for $T_{\pi}$ = 
142~MeV results from absorption, which is absent in 
our model approach for the amplitude of the $\pi A$ 
interaction.  Usually, the absorption on a nucleus 
is due to the reaction $\pi (NN)\to NN$, where $(NN)$ 
is a pair of correlated nucleons.  But the total 
cross section of the $\pi d\to NN$ reaction has its 
maximum at $T_{\pi}$ = 140~MeV, and the absorption 
cross section is somewhat suppressed at $T_{\pi}$ = 
180~MeV \cite{pdpwa}.  However, $T_{\pi}$ = 180~MeV 
corresponds to the maximum of the total ${\pi 
^{\pm}}d$ cross section, as shown in Fig.~\ref{f16}.  
The pion absorption cross section for $^3He$, 
measured at PSI \cite{psi}, also peaks at $T_{\pi}$ 
= 140 $-$ 150~MeV, as shown in Fig.~\ref{f16} as well.  
Therefore, if absorption were responsible for the 
suppression of the scattering cross sections at small 
scattering angles for $T_{\pi}$ = 142~MeV, then this 
effect would be considerably smaller for $T_{\pi}$ = 
180~MeV, where our fit to the data is much better.

\subsection{CSV in ${\pi ^{\pm}}{^4He}$ Elastic 
Scattering}
\label{sec:4He}

Data for the elastic $\pi ^4He$ differential cross 
sections have been obtained at LAMPF for $T_{\pi}$
below, at, and above the $\Delta _{33}$ resonance 
\cite{br91}.  The spin-flip amplitudes for elastic 
${\pi ^{\pm}}{^4He}$ in the single-scattering
approximation do not contribute to the $\Delta 
_{33}$ resonance.  Preliminary analysis of the 
charge asymmetry for $T_{\pi}$ = 180~MeV shows a 
statistically significant effect in $A_{\pi}$, of 
the order of 10\% at scattering angles $\theta\sim 
70^{\circ} - 90^{\circ}$ \cite{mor}.  The size of 
$A_{\pi}$ for $T_{\pi}\le 180~MeV$ and the 
possible influence of absorption there are still 
unknown.

Because of the larger role played by pion 
absorption on heavier nuclei, CSV is expected to be 
suppressed relative to the three- and four-body 
nuclei.  If present, CSV in heavier nuclei probably 
depends more on their geometrical properties than 
on the $\Delta _{33}$-mass splitting.  

\section{Summary}
\label{sec:conc}

We have performed theoretical calculations for the 
simple ratios $r_1$ and $r_2$ and the superratio $R$ 
for elastic ${\pi ^{\pm}}{^3H/^3He}$ scattering, for 
T$_{\pi}$ = 142, 180, 220, and 256~MeV and over a 
broad angular range.  We have found reasonable 
agreement between the results of our calculations 
and the experimental data, shown in Fig.~\ref{f1}, 
over most of the range of the data.

Our calculations were done in an approach 
utilizing the sum of the single- and 
double-scattering $\pi N$ contributions, as 
indicated in Fig.~\ref{f2}.  We took into account 
three sources of CSV $-$ the $\Delta _{33}$-mass
splitting and the external and internal Coulomb   
interactions.  We used $S$-shell WFs for $^3H$ 
and $^3He$.  This approach enabled us to use 
simple analytical expressions for the 
double-scattering contribution to pion-nuclear 
scattering, taking into account all spin and 
isospin amplitudes.  We used two different 
radial WFs for the $A = 3$ nuclei: \\

i) A simple Gaussian distribution (Eq.~(\ref{7}))
   with the slope describing the charge
   densities of $^3H$ and $^3He$ obtained from
   electron scattering 
   \cite{juster,mccarty,beck,amroun}.  We used
   the WF of Ref.~\cite{kam}. \\

ii) A sum of two Gaussian WFs~(Eq.~(\ref{8})),
    as used in Ref.~\cite{foursat} for the  
    description of the differential cross    
    sections of the reaction $^4He(p, d)     
    ^3He$.  This WF reproduces the minimum of  
    the $^3He$ charge form factor at $Q =    
    670~MeV/c$, but shows an enhancement at
    smaller momentum transfer.  \\

The calculated cross sections, shown in Figs.
~\ref{f3} $-$ \ref{f6}, agree qualitatively
with the experimental data.  For T$_{\pi}$ =
180~MeV, the theoretical curves have minima
at $\theta\sim 140^{\circ} - 150^{\circ}$
and reproduce the gradual growth of the cross
sections as $\theta$ approaches $180^{\circ}$,
indicating the importance of the inclusion of
double scattering for the differential cross 
sections.  Of course, the absolute cross
sections are very sensitive to the WF, and are 
not reproduced well by the simple $S$-shell 
approach used here.
   
The main goal of our study is the calculation
of the CSV effects for ${\pi ^{\pm}}{^3H/^3He}$ 
differential cross sections in terms of the
observables $r_1$, $r_2$, and $R$.  No free 
parameters are used in our approach in taking 
into account the $\Delta _{33}$-mass splitting 
and the external Coulomb interaction.  
Figures.~\ref{f8} $-$ \ref{f10} show that these 
factors alone account qualitatively for major 
features of the data.  These figures also show 
that there is little sensitivity to the choice 
of the wave function.

Figures~\ref{f12} $-$ \ref{f14} show that when 
the internal Coulomb effect is included as 
well, there is reasonable agreement between our 
theoretical calculations and the experimental 
data.  The best agreement is found for T$_{\pi}$ 
= 180~MeV (at the peak of the $\Delta 
_{33}$ resonance).  Both the $\Delta _{33}$-mass 
splitting and the internal Coulomb interaction 
are important for the reproduction of the shape 
of the angular distribution, both near the
non-spin-flip dip at $\theta\sim 80^{\circ}$ and at 
large scattering angles.  Although the influence 
of the internal Coulomb interaction on CSV has 
been shown before \cite{GG}, our investigation 
shows that including the $\Delta _{33}$-mass 
splitting results in a still better description 
of the effect of CSV, as it should: the $\Delta 
_{33}$-mass splitting exists, so its effects 
should not be ignored.
We also predict the simple mirror ratios for the
total cross sections, as shown in Fig.~\ref{f15}.

Finally, however, as seen from Figs.~\ref{f7} and 
\ref{f11}, our calculations do not reproduce the 
data for $T_{\pi}$ = 142~MeV.  Although we tried 
to take into account a number of different 
approaches beyond the framework of our model
(more accurate amplitudes for single and double 
scattering and Fermi motion, {\it {etc.}}), we 
were not able to improve the agreement for 
$T_{\pi}$ = 142~MeV.  Therefore, the question of 
the nature of the effect of CSV for $T_{\pi}$ = 
142~MeV remains open.  Perhaps there is an 
additional mechanism at $T_{\pi}$ = 142~MeV 
which does not manifest itself at higher 
energies; Fig.~\ref{f16} shows that quite 
possibly, pion absorption plays a major role.

\acknowledgments

The authors acknowledge very useful communications 
with B. V. Geshkenbein, S. S. Kamalov, and C. L. 
Morris.  This work was supported in part by the 
U.~S.~Department of Energy under Grants 
DE--FG02--95ER40901 and DE--FG02--99ER41110
and by the Russian Grants for Basic Research 
N~98--02--17618 and N~00--15--96562.  A.~K. 
acknowledges the hospitality extended by the 
Center for Nuclear Studies of The George Washington 
University.   I.~S. acknowledges partial support 
from Jefferson Lab, by the Southeastern Universities 
Research Association under DOE contract 
DE--AC05--84ER40150.

\appendix
\section{The Charge Form Factor F($\vec\Delta$)}
\label{sec:CFF}

Let us introduce relative coordinates of the 
nucleons ${\vec\rho}_i = {\vec r}_j - {\vec r}_k$ 
and ${\vec R}_i = \frac{1}{2}({\vec r}_j + {\vec r}
_k) - {\vec r}_i$ instead of $\vec r_i$.  In terms 
of these new variables, the function $\psi$ of 
Eq.~(\ref{5}) yields
\begin{eqnarray}
\psi (\vec{r}_1, \vec{r}_2, \vec{r}_3)\equiv\psi 
(\sum _{m = 1}^{3}(\vec r_i - {\vec R}_0)^2) = \psi 
(\frac{{\vec\rho}_i^2}{2} + \frac{2}{3}{\vec R}_i^2) 
= \psi (\vec{\rho}, \vec{R}).
\label{A1}\end{eqnarray}
This function $\psi$ does not depend on the 
selection of a basis $i(jk)$.  The Fourier 
transformation of the function $\psi$ is defined by 
\begin{eqnarray}
\varphi (\vec q, \vec Q) = \int d^3 \vec\rho d^3 
\vec R \psi(\vec\rho , \vec R) \exp(-i \vec 
q\cdot\vec\rho - i \vec Q\cdot\vec R),
\label{A2}\end{eqnarray}
where $\vec q = \vec q_i = (\vec p_j - \vec p_k)/2$ 
and $\vec Q = \vec Q_i = (\vec p_j + \vec p_k - 
2 \vec p_i)/3$ are relative momenta and ($\vec p_i$, 
$\vec p_j$, $\vec p_k$) are the momenta of the 
nucleons of the nucleus.  For the WF~(\ref{8}), 
$\varphi (\vec q, \vec Q)$ has the form
\begin{eqnarray}
\varphi (\vec q, \vec Q) = N (\frac{\pi}{\sqrt{12}}) 
\sum _{m}\frac{D_m}{\alpha _m ^3}\exp(- \frac{q^2}
{\alpha_m} - \frac{3 Q^2}{4 \alpha _m}),
\label{A3}\end{eqnarray}
where 
\begin{eqnarray}
N^{-2} = \frac{9}{2}(\pi \sqrt{12})^3 \sum
_{m,n}\frac{D_m~D_n}{(\alpha _m + \alpha _n)^2}.
\nonumber\end{eqnarray}
The charge form factor for elastic scattering is 
defined by
\begin{eqnarray}
F(\vec\Delta) = \frac{9}{2}\int \frac{d \vec 
q}{(2 \pi )^3}\frac{d \vec Q}{(2 \pi )^3} 
\varphi ( \vec q, \vec Q - \frac{2}{3} 
\vec\Delta) \varphi ( \vec q, \vec Q),~~
F(0) = 1.
\label{A4}\end{eqnarray}
An analytical expression for the form factor 
(\ref{A4}) corresponding to the WF~(\ref{8}) is
\begin{eqnarray}
F(\vec\Delta) = N_0^{-1} \sum _{m,n}\frac{D_m 
D_n}{(\alpha _m + \alpha _n)^3}\exp[- \frac{\Delta 
^2}{3 (\alpha _m + \alpha _n)}],
\label{A5}\end{eqnarray}
where
\begin{eqnarray}
N_0 = \sum _{m,n}\frac{D_m D_n}{(\alpha _m +
\alpha _n)^3}.
\nonumber\end{eqnarray}

\section{The Tensor $I_{\rm {ij}}$ and The Functions 
$J_1$ and $J_2$}
\label{sec:tensor}

The expression for the tensor $I_{ij}$ in terms 
of the wave function $\varphi ( \vec q, \vec Q$ ) 
(\ref{A2}) is
\begin{eqnarray}
I_{ij} = 4 \pi\int\frac{d^3 \vec q}{(2 \pi)^3}
\frac{d^3 \vec Q}{(2 \pi)^3}\frac{d^3 \vec 
Q^{\prime}}{(2 \pi)^3}~\varphi(\vec{q^{\prime}}, 
\vec{Q}^{\prime})~\varphi (\vec q, \vec Q)\frac
{\hat s_i \hat s_j}{k_1 ^2 - s^2 - i 0}.
\label{B1}\end{eqnarray}
This integral is suitable for calculations in 
coordinate space.  To do this, we follow a 
transformation first used in Ref.~\cite{aek1}:
\begin{eqnarray}
\frac{\hat s_i \hat s_j}{k_1 ^2 - s^2 - i 0} 
= \frac{1}{4 \pi}\int exp(i \vec s\cdot\vec r) 
H_{ij}(\vec r) d \vec r,
\label{B2}\end{eqnarray}
where
\begin{eqnarray}
H_{ij}(\vec r) = h_1 (r) \hat r_i \hat r_j + 
h_2 (r) \delta _{ij}
\nonumber\end{eqnarray}
and
\begin{eqnarray}
h_1 (r) = \frac{e^{i k r}}{r} +
\frac{3 i e^{i k r}}{k r^2} - 
\frac{3 e^{i k r}}{k^2 r^3} +
\frac{3}{k^2 r^3}, \cr
h_2 (r) = \frac{e^{i k r}}{k^2 r^3} -
\frac{1}{k^2 r^3} -
\frac{i e^{i k r}}{k r^2}.\phantom{xxxx}
\nonumber\end{eqnarray}
Then, using Eq.~(\ref{B2}) and the coordinate
expression for the wave function (\ref{A1}), we 
get the expression for the tensor $I_{ij}$:
\begin{eqnarray}
I_{ij} = \frac{2}{9}\int d^3 \vec\rho d^3 
\vec r \psi ^2 (\vec\rho, \vec R) H_{ij}
(\vec r) exp[i(\vec k_1 - \frac{\vec\Delta}
{3})\cdot\vec r + i \frac{\vec\Delta\cdot
\vec\rho}{3}],
\label{B3}\end{eqnarray}
where $\vec R = \vec r + \frac{1}{2}\vec
\rho$.  If the WF $\psi (\vec\rho, \vec R)$ is 
expressed as a sum of several Gaussians, the 
integral (\ref{B3}) can be represented in the 
form of expression (\ref{20}).  Then the 
integrals $J_1$ and $J_2$ are transformed into 
one-dimensional integrals.  For the WF~(\ref{8}), 
they have the form
\begin{eqnarray}
J_{1,2} = \frac{2}{9}N^2 \sum _{m,n}D_m D_n 
(\frac{3 \pi}{a_{mn}})^{1/2}\exp(- \frac
{\Delta ^2}{12 a_{mn}}) F_{1,2}(a_{mn}, 
\theta), \phantom{xx}\cr
~{\rm where}~
F_1(a, \theta) = \pi\int\limits_{0}^{\infty}
r^2 \exp(\frac{a r^2}{4}) (3 E_2 - E_0) h_1
(r) dr, \phantom{xxxxx}\cr
F_2(a, \theta) = \pi\int\limits_{0}^{\infty}
r^2 \exp(\frac{a r^2}{4}) [(E_0 - E_2) h_1(r) 
+ 2 E_0 h_2(r)] dr, \phantom{x}\cr
E_n = \int\limits_{-1}^{1}\exp(i \xi r z) 
z^n dz,\phantom{xxxxxxxxxxxx}
\label{B4}\end{eqnarray}
$\xi = k \cos\frac{\theta}{2}$, $\Delta = 
2 k \sin\frac{\Delta}{2}$, and $N$ is given by 
Eq.~(\ref{A3}).  In the case of WF~(\ref{7}), 
the expressions for the integrals $J_{1,2}$ are 
computed easily from Eq.~(\ref{B4}).

\eject

\eject

\begin{figure}[ht]
\centerline{
\psfig{file=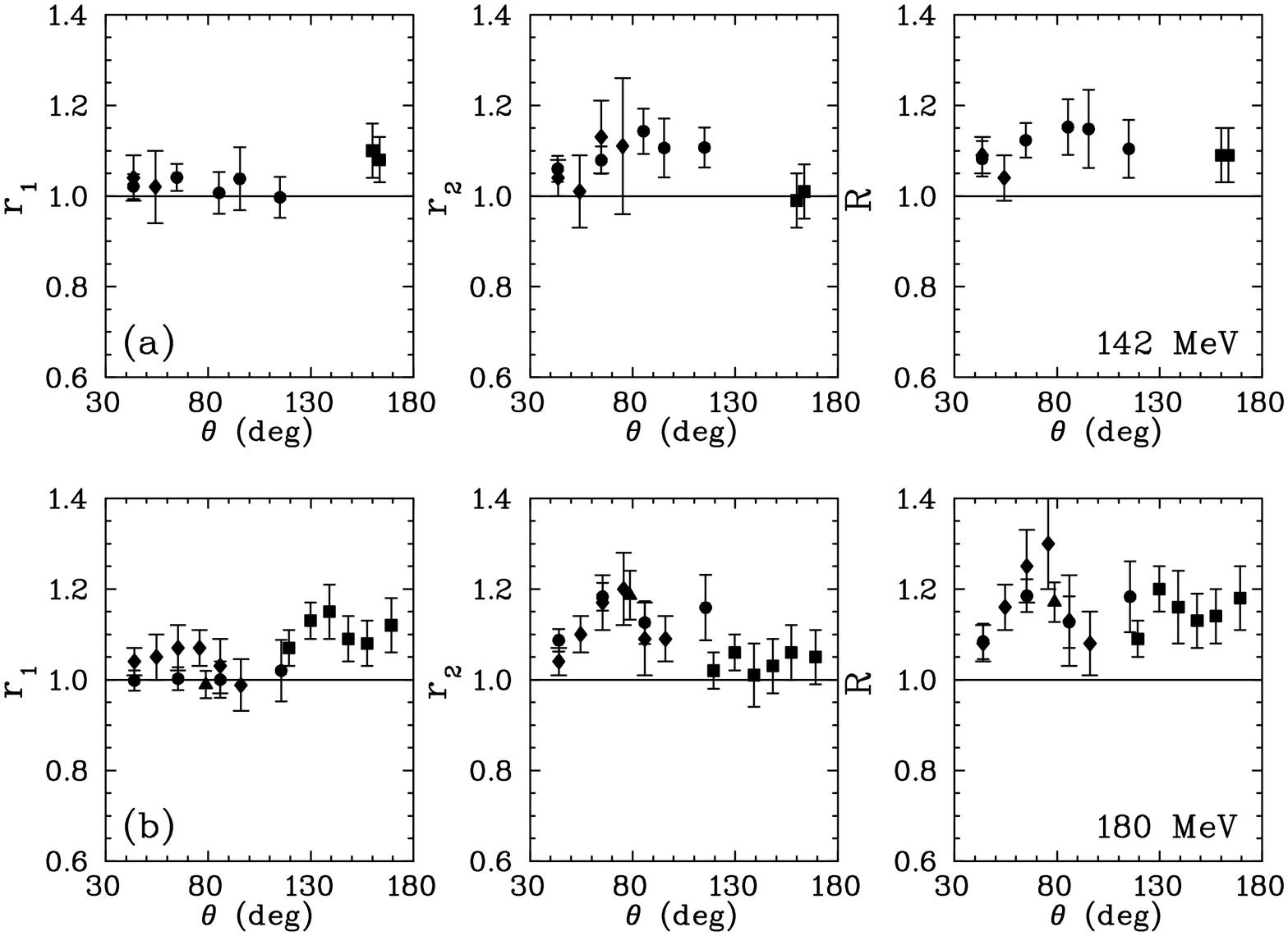,width=3in,clip=,silent=,angle=90}\hfill
\psfig{file=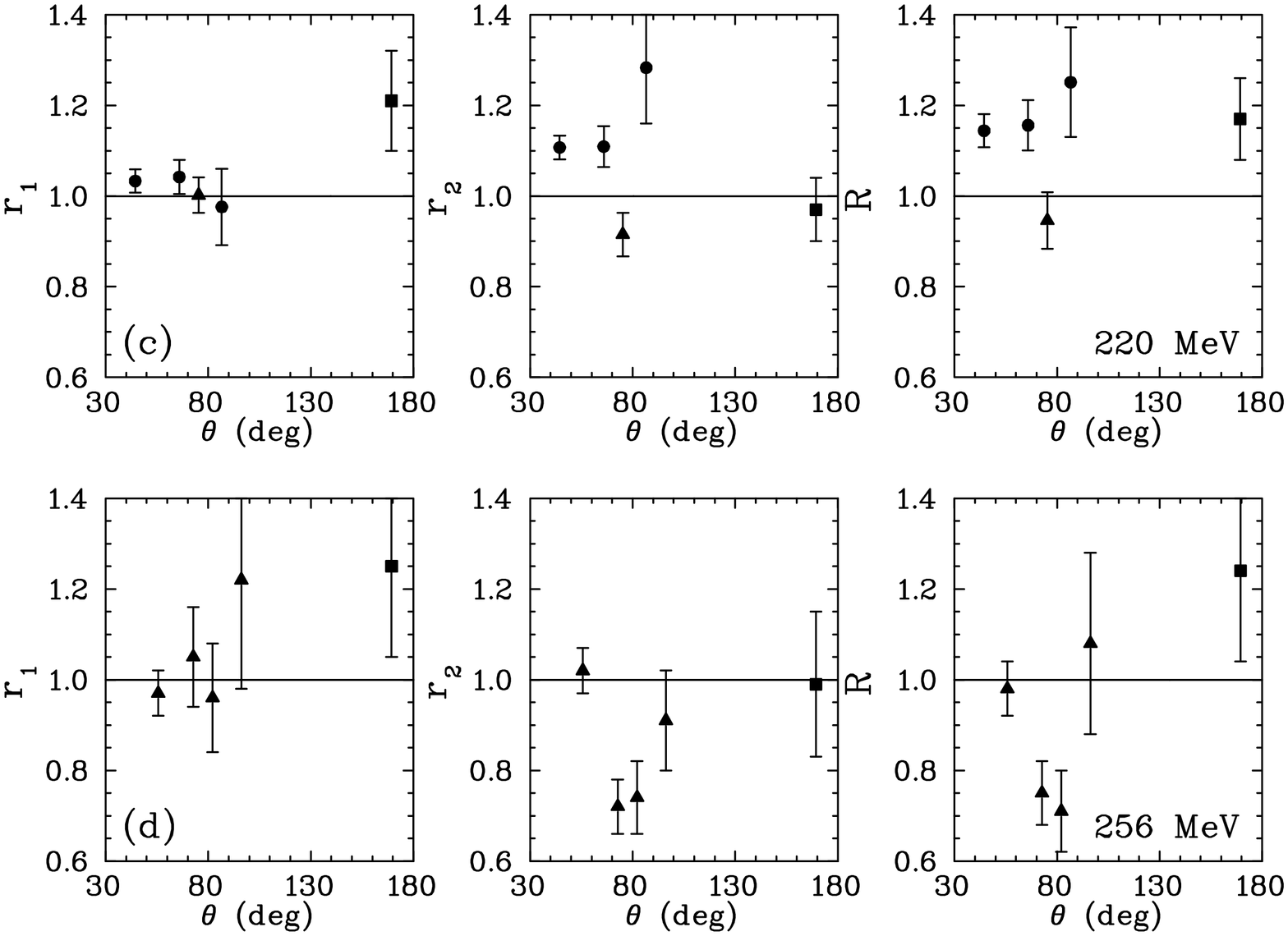,width=3in,clip=,silent=,angle=90}}
\vspace{3mm}
\caption[fig1]{\label{f1}
 The ratios $r_1$, $r_2$, and the superratio $R$
 for ${\pi ^{\pm}} {^3H/^3He}$ elastic scattering
 for various incident pion kinetic energies:  (a)
 T$_{\pi}$ = 142~MeV, (b) 180~MeV, (c) 220~MeV,
 and (d) 256~MeV.  Experimental data are from   
 \protect\cite{ne90} (diamonds),
 \protect\cite{pi91} (circles),
 \protect\cite{dh96} (triangles), and
 \protect\cite{ma00} (squares).}
\end{figure}

\begin{figure}[ht]
\centerline{
\psfig{file=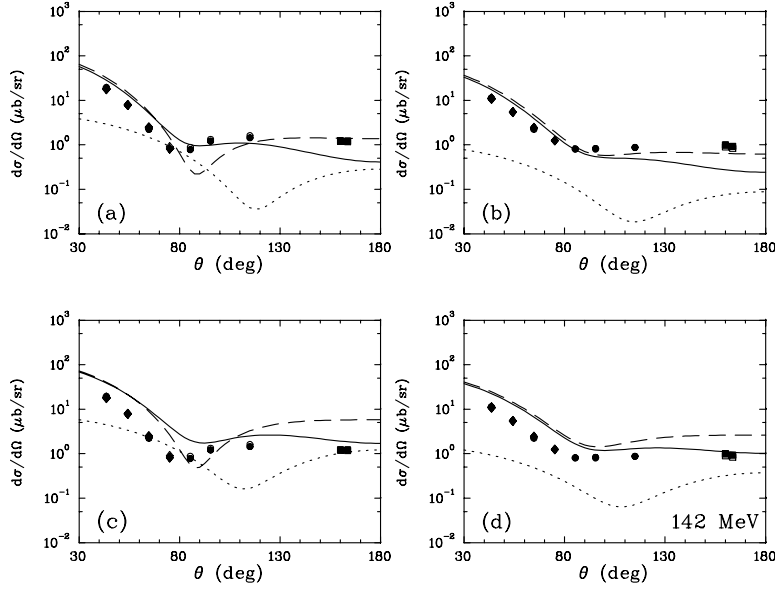,width=4in,clip=,silent=,angle=90}} 
\vspace{3mm}
\caption[fig3]{\label{f3}
 Differential cross sections for ${\pi ^{\pm}}
 {^3H/^3He}$ elastic scattering for $T_{\pi}$ = 
 142~MeV.  Plotted are results for (a,b) 
 WFs~(\ref{7}) and (c,d) WF~(\ref{8}).  Experimental
 data are from
 \protect\cite{ne90} (diamonds),
 \protect\cite{pi91} (circles), 
 \protect\cite{dh96} (triangles), and
 \protect\cite{ma00} (squares), with
 ${\pi ^+}{^3He/^3H}$ (filled) and
 ${\pi ^-}{^3He/^3H}$ (open).  The cross sections (a,c)
 are for ${\pi ^+}{^3He}$ and ${\pi ^-}{^3H}$ and (b,d)
 for ${\pi ^+}{^3H}$ and ${\pi ^-}{^3He}$.  The
 solid curves give the total contribution.  Results
 for single and double scattering alone are shown
 by the dashed and dotted curves, respectively.}
\end{figure}

\begin{figure}[ht]
\centerline{
\psfig{file=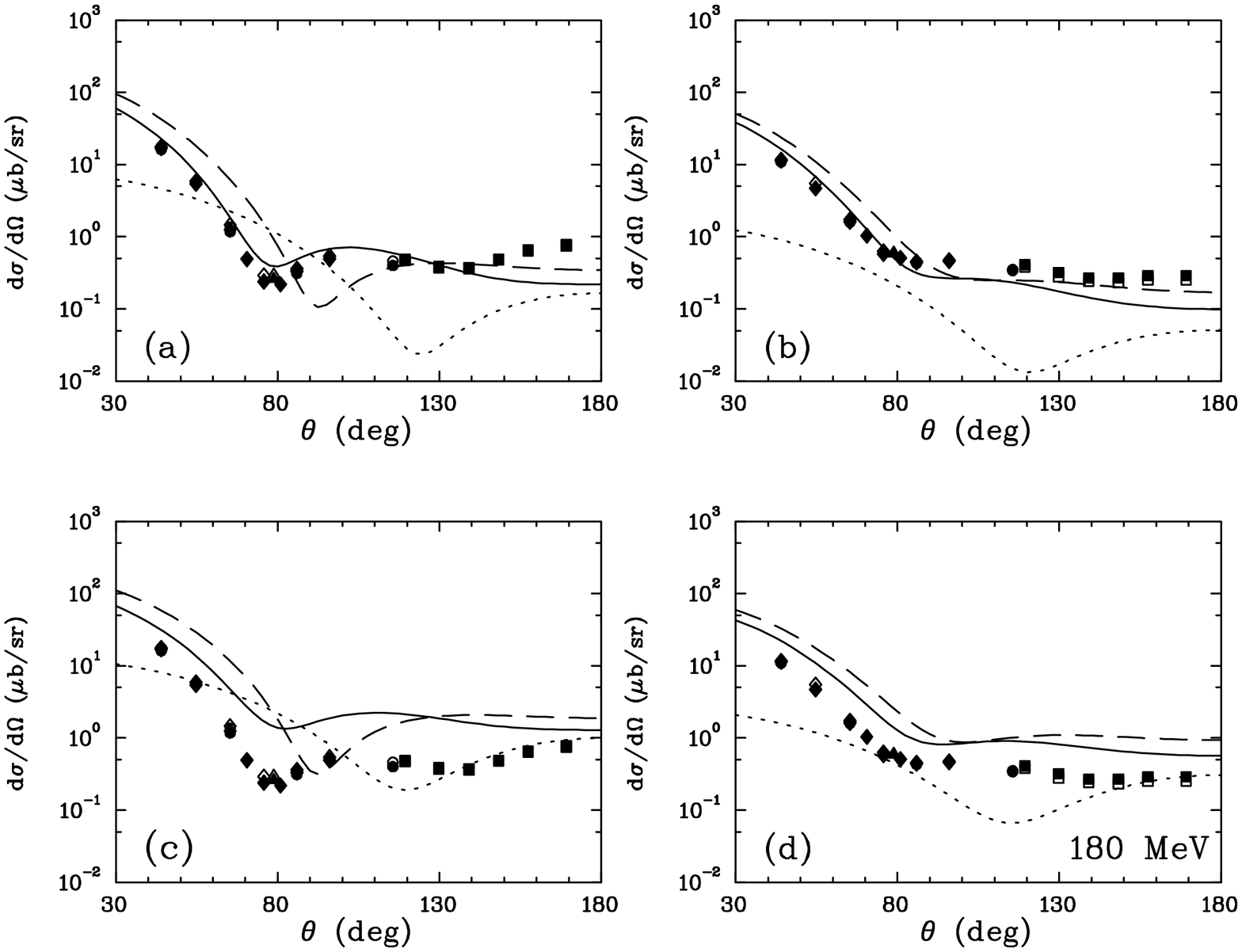,width=4in,clip=,silent=,angle=90}} 
\vspace{3mm}
\caption[fig4]{\label{f4}
 Differential cross sections for ${\pi^{\pm}}
 {^3H/^3He}$ elastic scattering for T$_{\pi}$ =
 180~MeV.  The notation is the same as for 
 Fig.~\ref{f3}.}
\end{figure}

\begin{figure}[ht]
\centerline{
\psfig{file=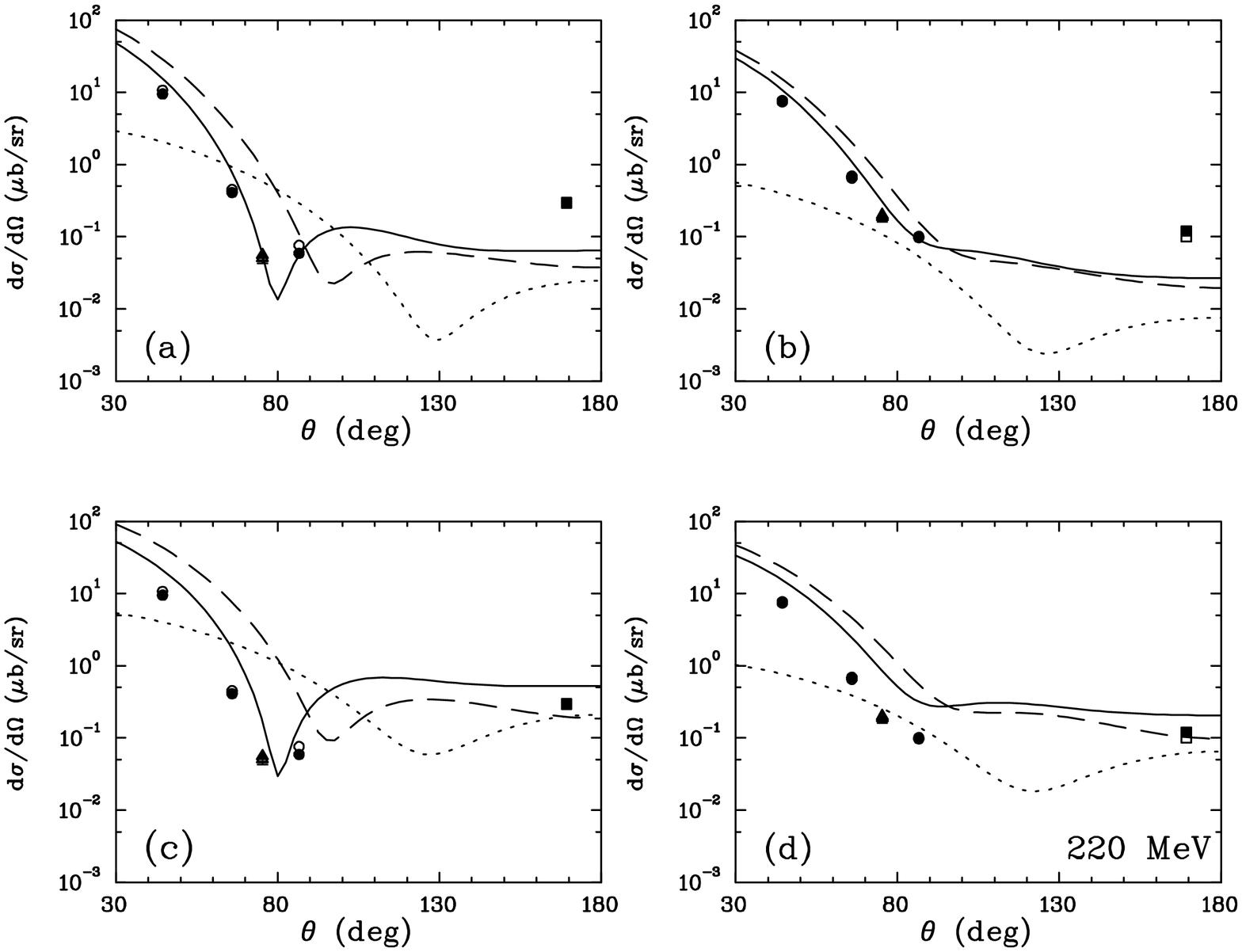,width=4in,clip=,silent=,angle=90}} 
\vspace{3mm}
\caption[fig5]{\label{f5}
 Differential cross sections for ${\pi^{\pm}}
 {^3H/^3He}$ elastic scattering for T$_{\pi}$ =
 220~MeV.  The notation is the same as for 
 Fig.~\ref{f3}.}
\end{figure}

\begin{figure}[ht]
\centerline{
\psfig{file=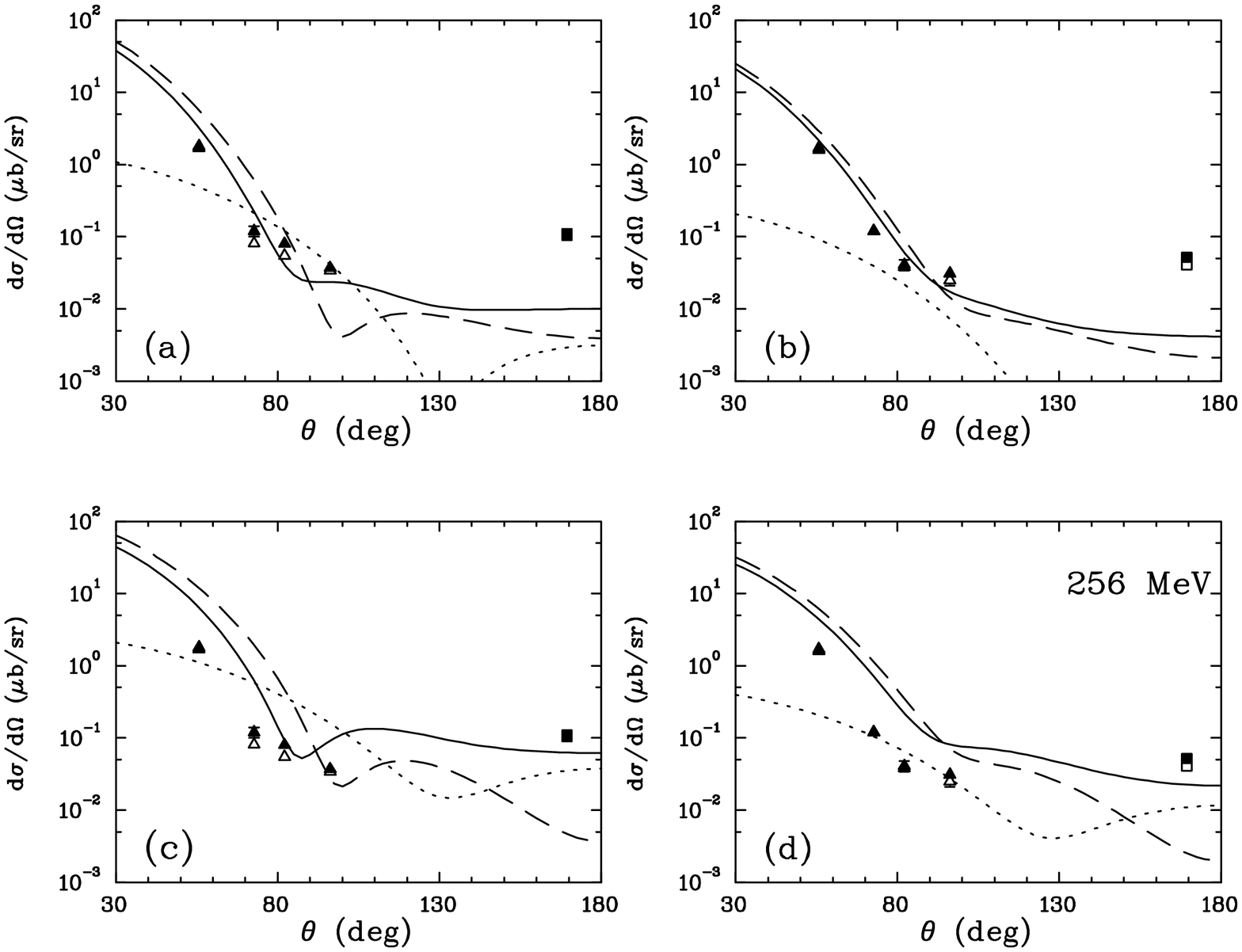,width=4in,clip=,silent=,angle=90}} 
\vspace{3mm}
\caption[fig6]{\label{f6}
 Differential cross sections for ${\pi^{\pm}}
 {^3H/^3He}$ elastic scattering for T$_{\pi}$ =
 256~MeV.  The notation is the same as for 
 Fig.~\ref{f3}.}
\end{figure}

\begin{figure}[ht]
\centerline{
\psfig{file=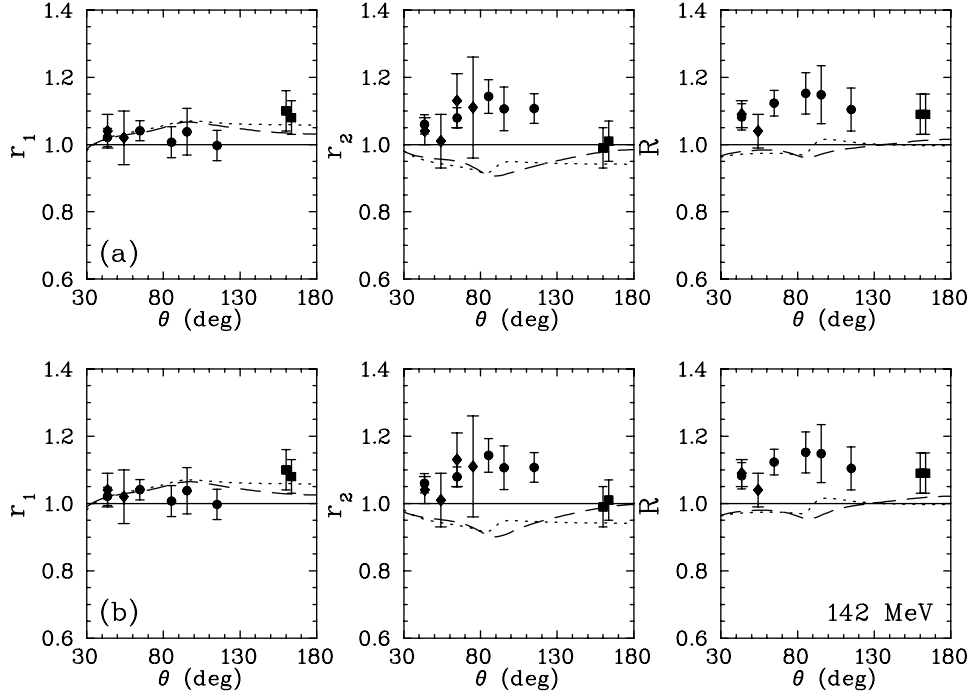,width=5in,clip=,silent=,angle=90}} 
\vspace{3mm}
\caption[fig7]{\label{f7}
 The ratios $r_1$ and $r_2$ and the superratio
 $R$ for T$_{\pi}$ = 142~MeV.  The notation for
 the experimental data is the same as for
 Fig.~\ref{f1}.  Only the $\Delta _{33}$-mass
 splitting and external Coulomb contributions
 are taken into account.  The full calculations
 take into account both single and double
 scattering, and are shown by the dashed curves. 
 The results for single scattering alone are
 shown by the dotted curves.  Plotted are the results
 for (a) WF~(\ref{7}) and (b) WF~(\ref{8}).}
\end{figure}

\begin{figure}[ht]
\centerline{
\psfig{file=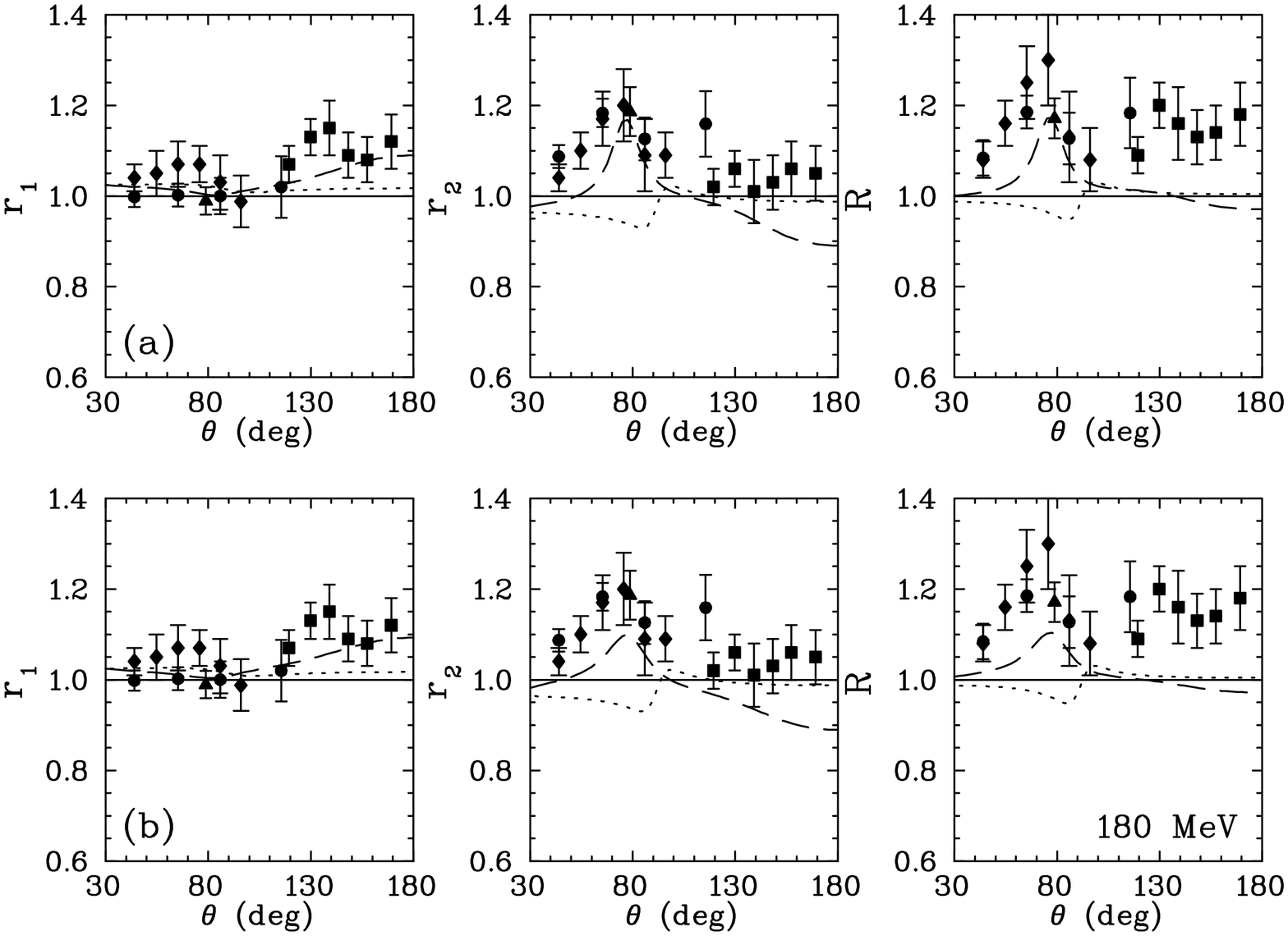,width=5in,clip=,silent=,angle=90}} 
\vspace{3mm}
\caption[fig8]{\label{f8}
 The ratios $r_1$ and $r_2$ and the superratio $R$
 for T$_{\pi}$ = 180~MeV.  The notation is the
 same as for Fig.~\ref{f7}.}
\end{figure}

\begin{figure}[ht]
\centerline{
\psfig{file=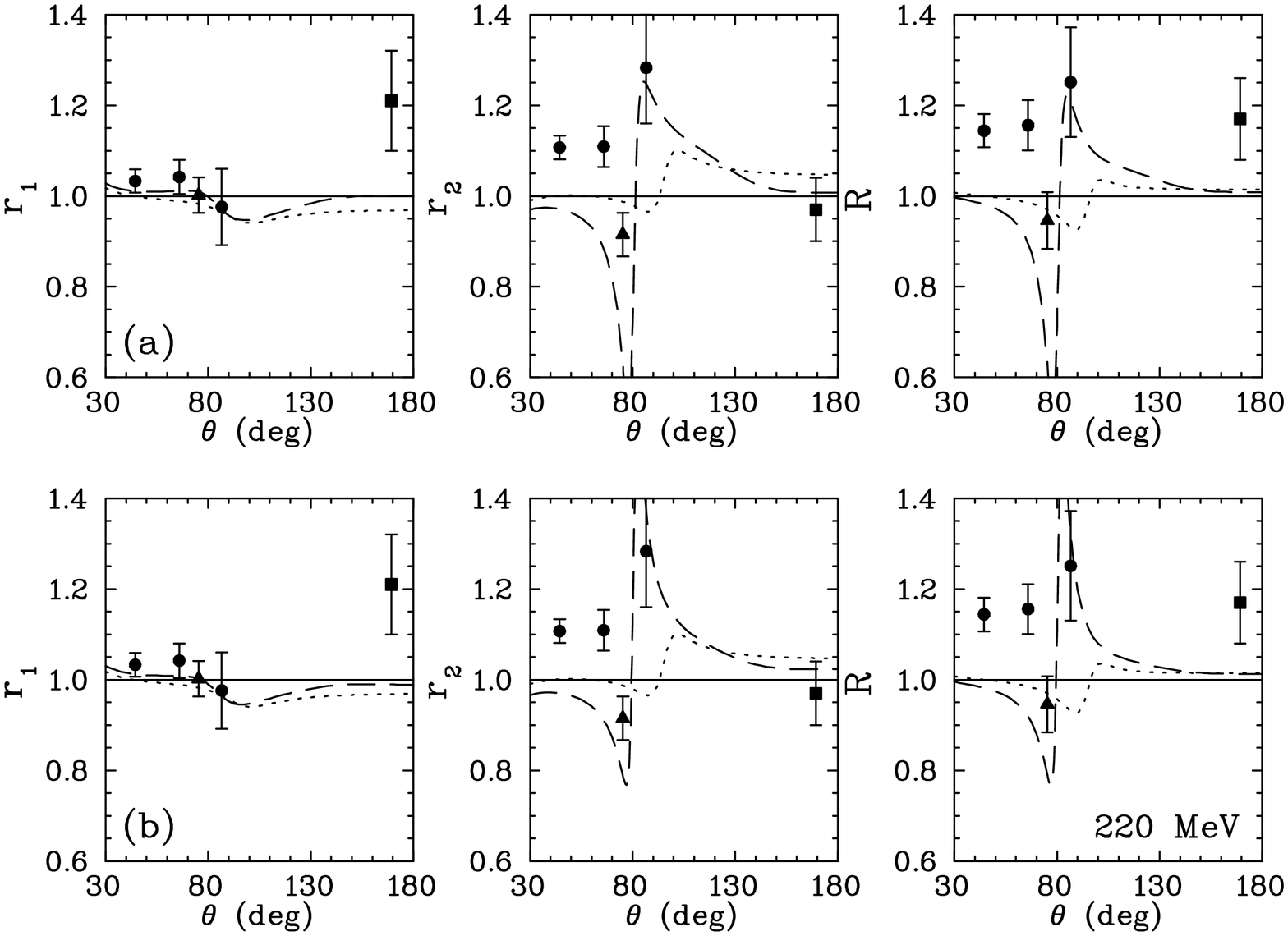,width=5in,clip=,silent=,angle=90}} 
\vspace{3mm}
\caption[fig9]{\label{f9}
 The ratios $r_1$ and $r_2$ and the superratio $R$
 for T$_{\pi}$ = 220~MeV.  The notation is the
 same as for Fig.~\ref{f7}.}
\end{figure}

\begin{figure}[ht]
\centerline{
\psfig{file=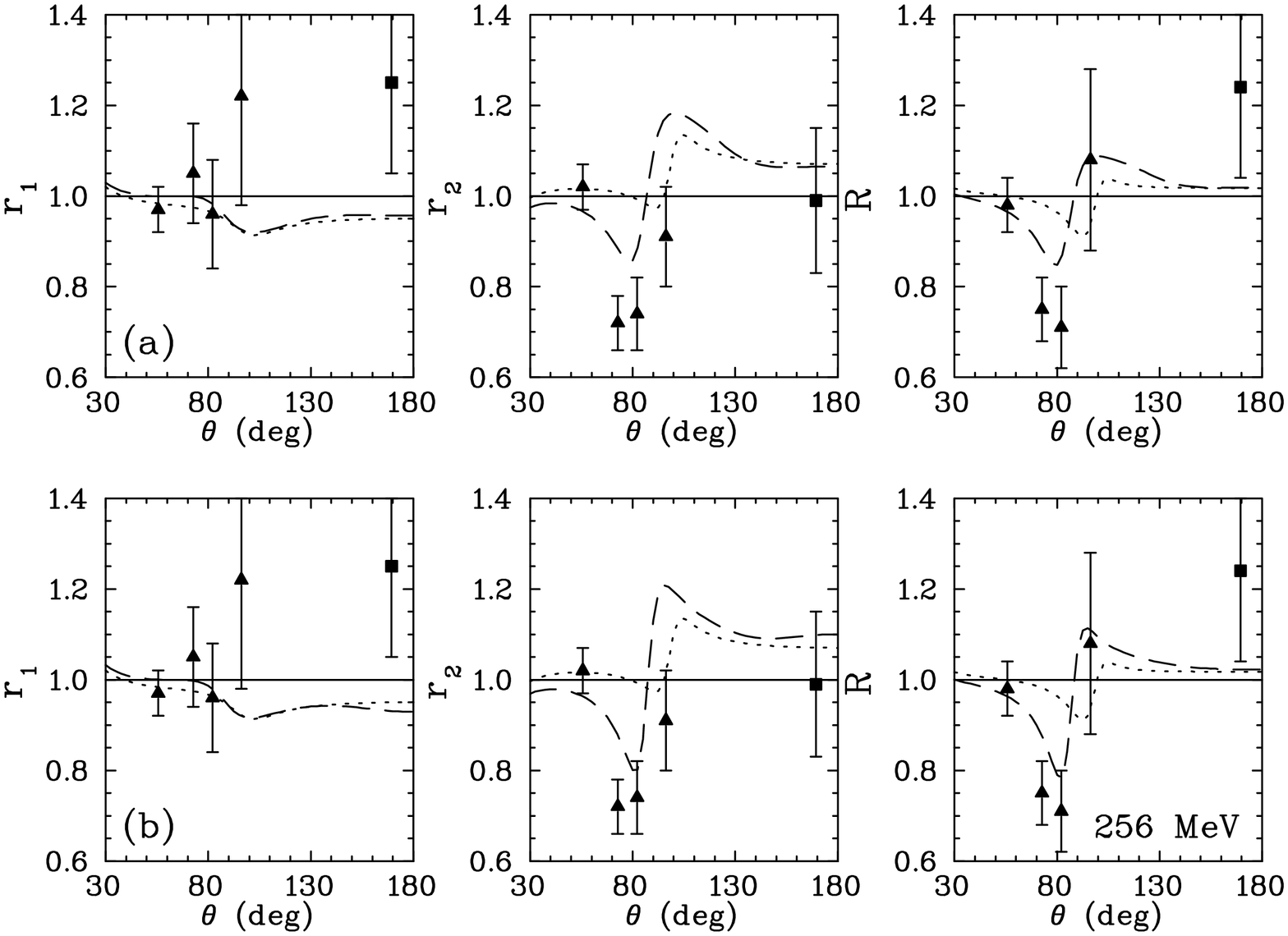,width=5in,clip=,silent=,angle=90}} 
\vspace{3mm}
\caption[fig10]{\label{f10}
 The ratios $r_1$ and $r_2$ and the superratio $R$
 for T$_{\pi}$ = 256~MeV.  The notation is the
 same as for Fig.~\ref{f7}.}
\end{figure}

\begin{figure}[ht]
\centerline{
\psfig{file=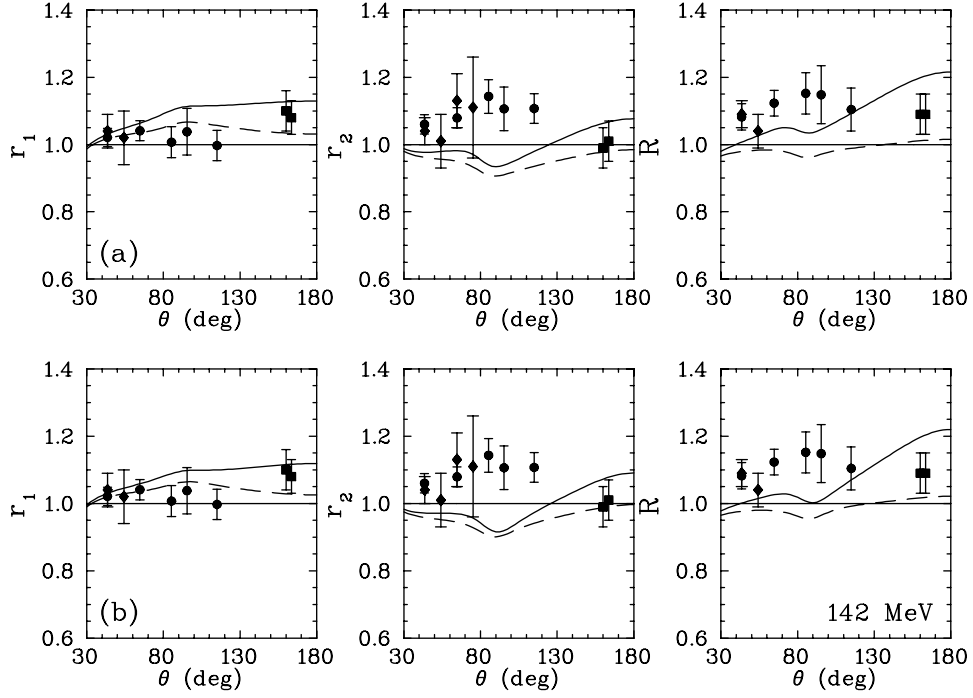,width=5in,clip=,silent=,angle=90}} 
\vspace{3mm}
\caption[fig11]{\label{f11}
 The ratios $r_1$ and $r_2$ and the superratio    
 $R$ for T$_{\pi}$ = 142~MeV.  The notation   
 for the experimental data is the same as for
 Fig.~\ref{f1}.  The $\Delta _{33}$-mass
 splitting with external (and internal)
 Coulomb contributions are shown by the dashed
 (solid) curves.  Plotted are the results for (a)
 WF~(\ref{7}) and (b) WF~(\ref{8}).  Version (iii)
 of the variation of the WFs is shown, as described
 in Sec.~\ref{sec:theor2}.}
\end{figure}

\begin{figure}[ht]
\centerline{
\psfig{file=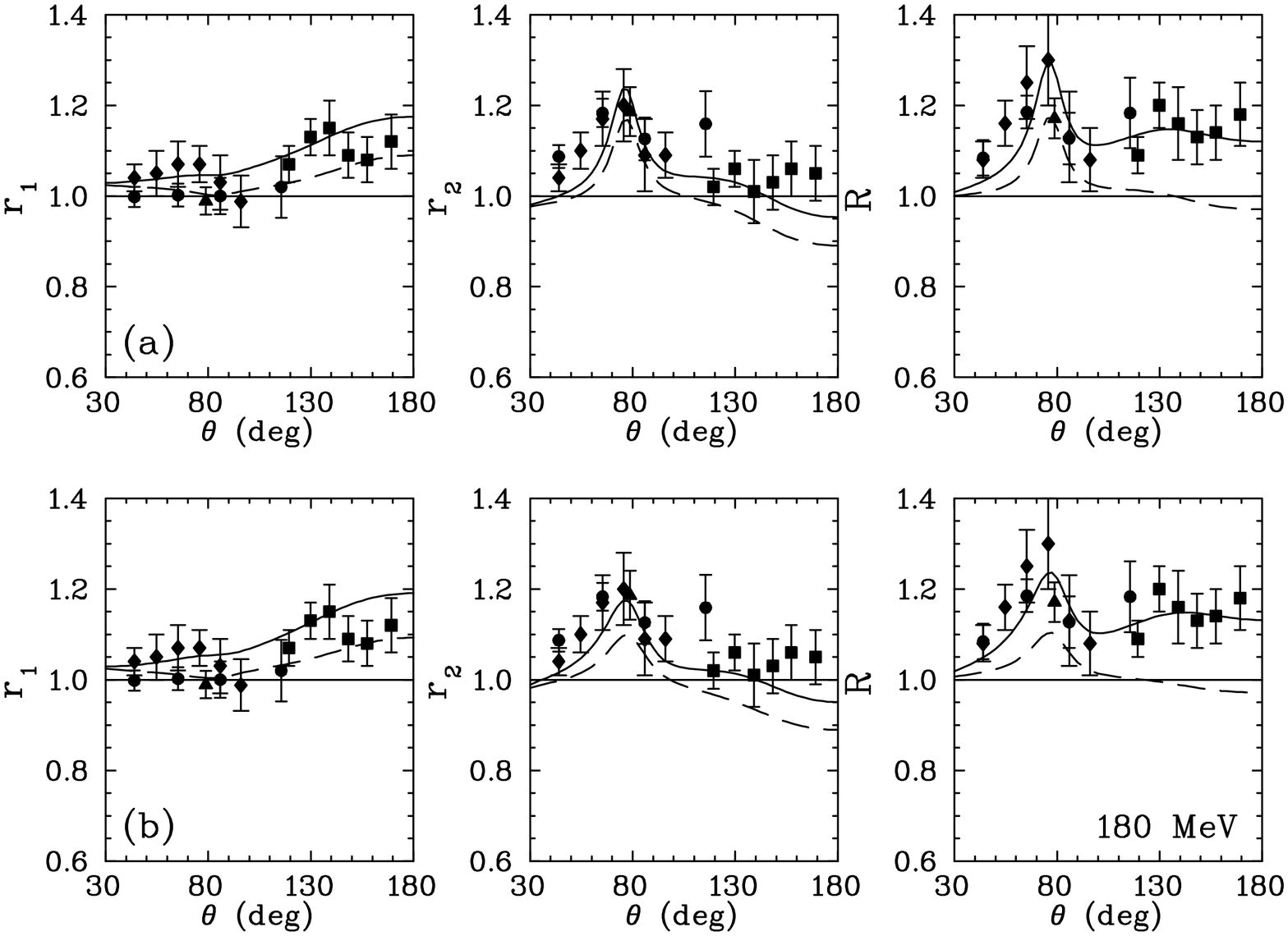,width=5in,clip=,silent=,angle=90}} 
\vspace{3mm}
\caption[fig12]{\label{f12}
 The ratios $r_1$ and $r_2$ and the superratio $R$
 for T$_{\pi}$ = 180~MeV.  The notation is the
 same as for Fig.~\ref{f11}.}
\end{figure}

\begin{figure}[ht]
\centerline{
\psfig{file=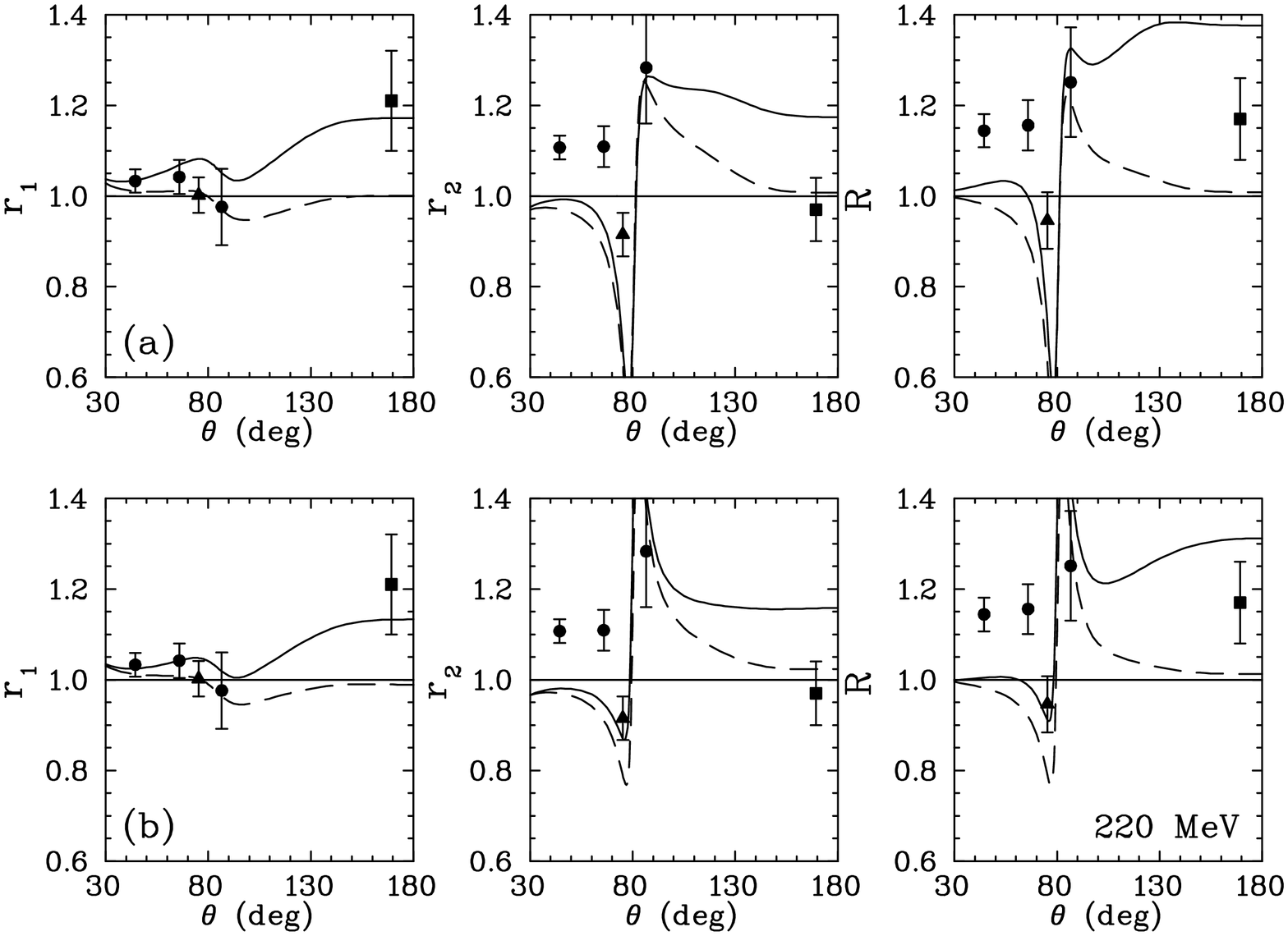,width=5in,clip=,silent=,angle=90}} 
\vspace{3mm}
\caption[fig13]{\label{f13}
 The ratios $r_1$ and $r_2$ and the superratio $R$
 for T$_{\pi}$ = 220~MeV.  The notation is the
 same as for Fig.~\ref{f11}.}
\end{figure}

\begin{figure}[ht]
\centerline{
\psfig{file=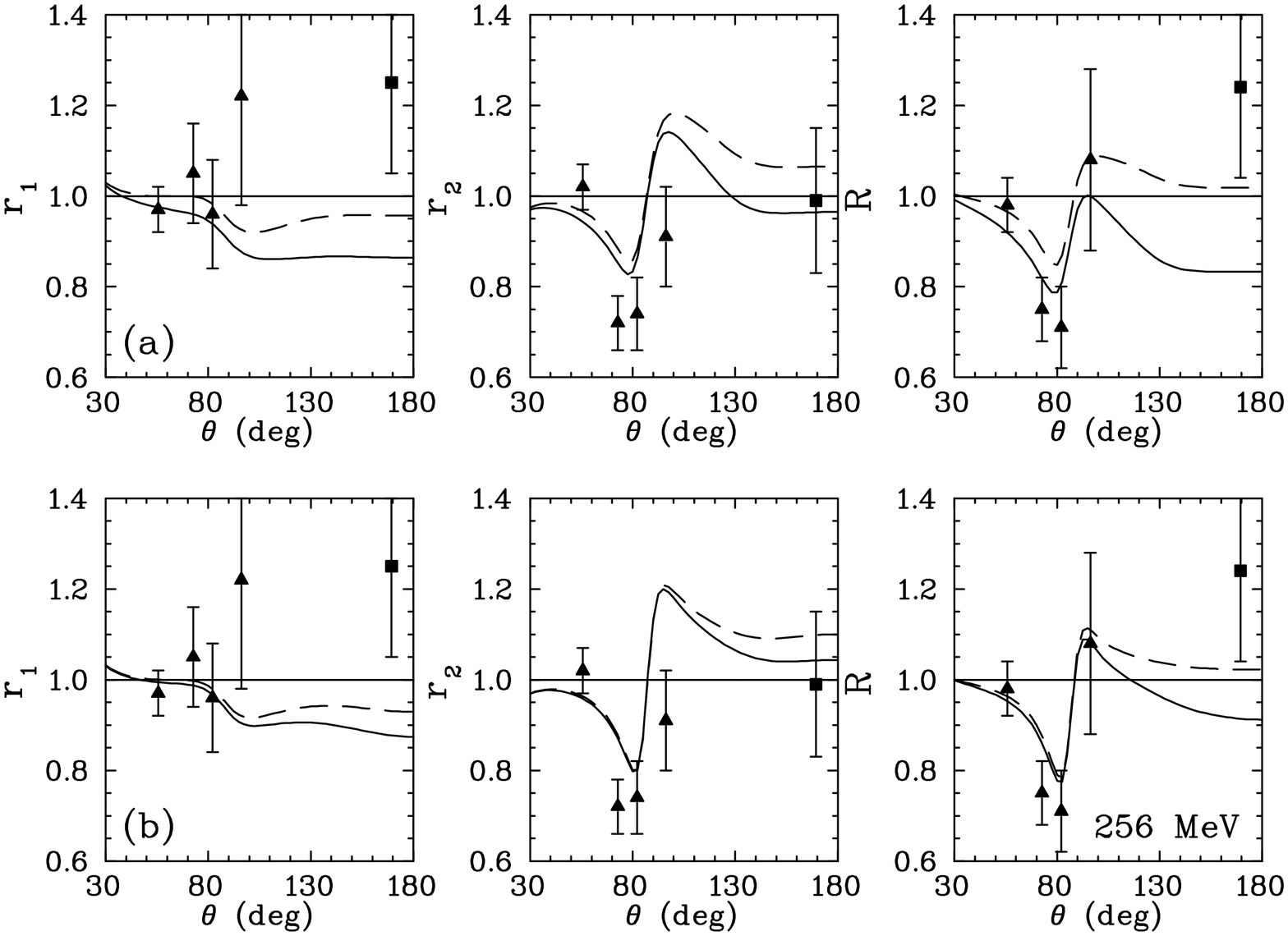,width=5in,clip=,silent=,angle=90}} 
\vspace{3mm}
\caption[fig14]{\label{f14}
 The ratios $r_1$ and $r_2$ and the superratio $R$
 for T$_{\pi}$ = 256~MeV.  The notation is the
 same as for Fig.~\ref{f11}.}
\end{figure}

\begin{table}[t]
\caption{Best-fit results for the ratios $r_1$ and 
         $r_2$ from variation of the $^3H$ and $^3He$ 
         WFs.  The slope $b$ has been varied for the 
         WF~(\ref{7}) \protect\cite{kam}.  Cases (i) 
         $-$ (iii) of variations of the WF~(\ref{8}) 
         \protect\cite{foursat} have been discussed in 
         the text.  Differences in the mean-square 
         charge radii $\delta r = r(^3{\rm He}) - 
         r(^3{\rm H})$ and $\chi ^2/d.f.$ are listed 
         in columns 5 and 6.  The last column shows 
         $\chi ^2/d.f.$ when the internal Coulomb 
         interaction is not taken into account.}
\label{tbl1}
\begin{center}
\begin{tabular}{|c|c|c|c|c|c|c|}
\rule{0pt}{16pt} 
$T_{\pi}$ (MeV) &
WF & 
Varied Parameter & &
$\delta r$ (fm) & 
$\chi^2/d.f.$ & 
$\chi^2/d.f.$ ($\delta r\!\equiv\! 0\,$) \\
\hline
\rule{0pt}{16pt} & \protect\cite{kam} & b & & 0.015 & 3.90 & 4.87 \\
\cline{2-7}
  142 &  & $\alpha_1$ & (i) & 0.017 & 4.15  &  \\
\cline{3-6}
  & \protect\cite{foursat} & $\alpha_2$ & (ii) & 0.010 & 4.44 & 5.06 \\
\cline{3-6}
  &  & $\alpha_1$ and $\alpha_2$ & (iii) & 0.014 & 4.26 &  \\
\hline
\rule{0pt}{16pt} & \protect\cite{kam} & b & & 0.012 & 1.68 & 2.88 \\
\cline{2-7}
  180  &  & $\alpha_1$  & (i) & 0.017 & 1.60 & \\
\cline{3-6}
  & \protect\cite{foursat} & $\alpha_2$ & (ii) & 0.014 & 1.65 & 3.05 \\
\cline{3-6}
  &  & $\alpha_1$ and $\alpha_2$ & (iii) & 0.016 & 1.61 &  \\
\hline
\rule{0pt}{16pt} & \protect\cite{kam} & b & & 0.019 & 8.46 & 10.5 \\
\cline{2-7}
  220 &  & $\alpha_1$  & (i) & 0.016 & 5.37 & \\
\cline{3-6}
  & \protect\cite{foursat} & $\alpha_2$ & (ii) & 0.011 & 5.37 & 6.74 \\
\cline{3-6}
  &  & $\alpha_1$ and $\alpha_2$ & (iii) & 0.014 & 5.36 &  \\
\hline
\rule{0pt}{16pt} & \protect\cite{kam} & b & & -0.010 & 2.26 & 2.43 \\
\cline{2-7}
  256 &  & $\alpha_1$  & (i) & -0.006 & 2.29 & \\
\cline{3-6}
  & \protect\cite{foursat} & $\alpha_2$ & (ii) & 0  & 2.37  & 2.37 \\
\cline{3-6}
  &  & $\alpha_1$ and $\alpha_2$ & (iii) & -0.003 & 2.35 &  \\
\end{tabular}
\end{center}
\end{table}  
{\Large\bf Figure captions} \\
\newcounter{fig}
\begin{list}
{Figure \arabic{fig}.}
{\usecounter{fig}\setlength{\rightmargin}{\leftmargin}}
\item
{The ratios $r_1$, $r_2$, and the superratio $R$ for
 ${\pi ^{\pm}} {^3H/^3He}$ elastic scattering for
 various incident pion kinetic energies: (a) T$_{\pi}$
 = 142~MeV, (b) 180~MeV, (c) 220~MeV, and (d) 256~MeV.  
 Experimental data are from
 \protect\cite{ne90} (diamonds),
 \protect\cite{pi91} (circles), 
 \protect\cite{dh96} (triangles), and
 \protect\cite{ma00} (squares).}
\item
{(a) Single- and (b) double-scattering diagrams 
 used in the present calculations for ${\pi ^{\pm}}
 {^3H/^3He}$ elastic scattering.}
\item
{Differential cross sections for ${\pi ^{\pm}}{^3H/
 ^3He}$ elastic scattering for $T_{\pi}$ = 142~MeV.  
 Plotted are results for (a,b) WFs~(\ref{7}) and
 (c,d) WF~(\ref{8}).  Experimental data are from
 \protect\cite{ne90} (diamonds),
 \protect\cite{pi91} (circles),
 \protect\cite{dh96} (triangles), and
 \protect\cite{ma00} (squares), with
 ${\pi ^+}{^3He/^3H}$ (filled) and 
 ${\pi ^-}{^3He/^3H}$ (open).  The cross sections 
 (a,c) are for ${\pi ^+}{^3He}$ and ${\pi ^-}{^3H}$
 and (b,d) for ${\pi ^+}{^3H}$ and ${\pi ^-}{^3He}$. 
 The solid curves give the total contribution. 
 Results for single and double scattering alone are 
 shown by the dashed and dotted curves, respectively.}
\item
{Differential cross sections for ${\pi^{\pm}}
 {^3H/^3He}$ elastic scattering for T$_{\pi}$ = 
 180~MeV.  The notation is the same as for 
 Fig.~\ref{f3}.}
\item
{Differential cross sections for ${\pi^{\pm}}
 {^3H/^3He}$ elastic scattering for T$_{\pi}$ =
 220~MeV.  The notation is the same as for 
 Fig.~\ref{f3}.}
\item
{Differential cross sections for ${\pi^{\pm}}
 {^3H/^3He}$ elastic scattering for T$_{\pi}$ =
 256~MeV.  The notation is the same as for 
 Fig.~\ref{f3}.}
\item
{The ratios $r_1$ and $r_2$ and the superratio
 $R$ for T$_{\pi}$ = 142~MeV.  The notation for
 the experimental data is the same as for
 Fig.~\ref{f1}.  Only the $\Delta _{33}$-mass
 splitting and external Coulomb contributions
 are taken into account.  The full calculations
 take into account both single and double
 scattering, and are shown by the dashed curves. 
 The results for single scattering alone are
 shown by the dotted curves.  Plotted are the 
 results for (a) WF~(\ref{7}) and (b)
 WF~(\ref{8}).}
\item
{The ratios $r_1$ and $r_2$ and the superratio $R$
 for T$_{\pi}$ = 180~MeV.  The notation is the
 same as for Fig.~\ref{f7}.}
\item
{The ratios $r_1$ and $r_2$ and the superratio $R$
 for T$_{\pi}$ = 220~MeV.  The notation is the
 same as for Fig.~\ref{f7}.}
\item
{The ratios $r_1$ and $r_2$ and the superratio $R$
 for T$_{\pi}$ = 256~MeV.  The notation is the
 same as for Fig.~\ref{f7}.}
\item
{The ratios $r_1$ and $r_2$ and the superratio
 $R$ for T$_{\pi}$ = 142~MeV.  The notation
 for the experimental data is the same as for
 Fig.~\ref{f1}.  The $\Delta _{33}$-mass
 splitting with external (and internal)
 Coulomb contributions are shown by the dashed
 (solid) curves.  Plotted are the results for (a)
 WF~(\ref{7}) and (b) WF~(\ref{8}).  Version (iii)
 of the variation of the WFs is shown, as
 described in Sec.~\ref{sec:theor2}.} 
\item
{The ratios $r_1$ and $r_2$ and the superratio $R$
 for T$_{\pi}$ = 180~MeV.  The notation is the
 same as for Fig.~\ref{f11}.}
\item
{The ratios $r_1$ and $r_2$ and the superratio $R$
 for T$_{\pi}$ = 220~MeV.  The notation is the
 same as for Fig.~\ref{f11}.}
\item
{The ratios $r_1$ and $r_2$ and the superratio $R$
 for T$_{\pi}$ = 256~MeV.  The notation is the
 same as for Fig.~\ref{f11}.}
\item
{Predictions for the ratios $r_{1t}$ (solid), and
 $r_{2t}$ (dashed) for the total ${\pi ^{\pm}}
 {^3H/^3He}$ cross sections.  Calculations were
 done for single and double scattering with the
 $\Delta _{33}$-mass splitting.  The Coulomb
 interactions are not taken into account.  Plotted
 are the results for (a) WF~(\ref{7}) and (b)
 WF~(\ref{8}).}
\item
{Total $\pi ^+$ cross sections.  Plotted are cross
 sections for $\pi ^+d$ (solid) and $\pi ^+d\to pp$
 (multiplied by a factor of 20) (dashed) from a
 recent combined fit of the $pp$ and $\pi d$ elastic
 scattering with the $\pi ^+d\to pp$ data
 \protect\cite{pdpwa}.  The $\pi ^3He$ absorption
 data (multiplied by a factor of 10) are from 
 \protect\cite{psi} (filled circles); it can be
 seen that they peak near T$_{\pi}$ = 140~MeV.}
\end{list}
\eject
\end{document}